\documentclass{nature_sjs}
\usepackage{aas_macros}
\usepackage{graphicx}%
\usepackage{multirow}%
\usepackage{amsmath,amssymb,amsfonts}%
\usepackage{amsthm}%
\usepackage{mathrsfs}%
\usepackage[title]{appendix}%
\usepackage{xcolor}%
\usepackage{textcomp}%
\usepackage{manyfoot}%
\usepackage{booktabs}%
\usepackage{algorithm}%
\usepackage{algorithmicx}%
\usepackage{algpseudocode}%
\usepackage{listings}%
\usepackage{ulem}
\usepackage{lscape}
\usepackage{units}
\usepackage{rotating}
\usepackage{hyperref}
\usepackage{tablefootnote}
\usepackage{amssymb,amsmath,caption}

\newcommand{\Msolar}{M_{\odot}}
\newcommand{\frb}{{FRB\,20201124A}}

\newcommand*\fdg{\ensuremath{\overset{^\circ}{.}}}

\newcommand*\farcs{\ensuremath{\overset{\prime\prime}{.}}}

\def\be{\begin{eqnarray}}
\def\ee{\end{eqnarray}}

\begin{document}
%\begin{linenumbers}
\title{A nebular origin for the persistent radio emission of fast radio bursts}
\author{\noindent
Gabriele Bruni$^{1}$,
Luigi Piro${^1}$, 
Yuan-Pei Yang$^{2,3}$, 
Salvatore Quai$^{4,5}$, 
Bing Zhang$^{6,7}$, 
Eliana Palazzi$^{5}$, 
Luciano Nicastro$^{5}$, 
Chiara Feruglio$^{8,9}$, 
Roberta Tripodi$^{10,8,9}$, 
Brendan O’Connor$^{11}$, 
Angela Gardini$^{12}$, 
Sandra Savaglio$^{13,5,14}$, 
Andrea Rossi$^{5}$, 
Ana M. Nicuesa Guelbenzu$^{15}$, 
Rosita Paladino$^{16}$
}

\maketitle

\begin{affiliations}
\item IAPS -- Institute for Space Astrophysics and Planetology, INAF, via del Fosso del Cavaliere, 100, Rome, 00133, Italy. 
\item South-Western Institute for Astronomy Research, Yunnan University, University City, Chenggong District, Kunming, 650504, Yunnan, China. 
\item Purple Mountain Observatory, Chinese Academy of Sciences, 10 Yuanhua Road, Qixia District, Nanjing, 210023, Jiangsu, China. 
\item Dipartimento di Fisica e Astronomia ‘Augusto Righi’, Università degli Studi di Bologna, UNIBO, via Piero Gobetti 93/3, Bologna, 40129, Italy. 
\item Osservatorio di Astrofisica e Scienza dello Spazio di Bologna, INAF, via Piero Gobetti 93/3, Bologna, 40129, Italy.
\item Nevada Center for Astrophysics, University of Nevada, NV 89154-4002, Las Vegas, Nevada USA.
\item Department of Physics and Astronomy, University of Nevada, NV 89154-4002, Las Vegas, Nevada USA.
\item Osservatorio Astronomico di Trieste, INAF, via G.B. Tiepolo 11, Trieste, 34143, Italy.
\item IFPU–Institute for Fundamental Physics of the Universe, via Beirut 2, Trieste, 34151, Italy.
\item Dipartimento di Fisica, Università di Trieste, via G.B. Tiepolo 11, Trieste, 34143, Italy.
\item McWilliams Center for Cosmology, Department of Physics, Carnegie Mellon University, 4000 Forbes Avenue, Pittsburgh, PA 15213, USA. 
\item Instituto de Astrof\'isica de Andaluc\'ia, CSIC, Glorieta de la Astronom\'ia s/n, Granada, 18008, Spain.
\item Dipartimento di Fisica, Università della Calabria, Arcavacata di Rende, 87036, Italy.
\item  Laboratori Nazionali di Frascati, INFN, Frascati, 00040, Italy
\item Th\"uringer Landessternwarte Tautenburg, Sternwarte 5, Tautenburg, 35 07778, Germany.
\item IRA – Istituto di Radioastronomia, INAF, via Piero Gobetti, 101, Bologna, 40129, Italy.

\end{affiliations}
\bigskip


{\abstract{Fast radio bursts (FRBs) are millisecond-duration, bright ($\sim$Jy) extragalactic bursts, whose production mechanism is still unclear \cite{2023RvMP...95c5005Z}. Recently, two repeating FRBs were found to have a physically associated persistent radio source of non-thermal origin \cite{michilli18,niu22}. These two FRBs have unusually large Faraday rotation measure values \cite{michilli18,niu22}, likely tracing a dense magneto-ionic medium, consistent with synchrotron radiation originating from a nebula surrounding the FRB source \cite{margalit18,metzger19,Yang2020,yang22,2022ApJ...937....5S}. Recent theoretical arguments predict that, if the observed Faraday rotation measure mostly arises from the persistent radio source region, there should be a simple relation between the luminosity of the latter and the first. \cite{yang20,yang22}. %\lp{addrf}. %Since low RM values correspond to low radio luminosities, this might explain why most of FRB discovered so far do not show an associated PRS. %The only two PRSs known so far fell on the predicted slope, but were not spaced enough to definitely assess the proposed RM vs Luminosity relation. 
We report here the detection of a third, less luminous persistent radio source associated with the repeating FRB source \frb\ at a distance of 413 Mpc, significantly expanding the predicted relation into the low luminosity --  low Faraday rotation measure regime ($<$1000 rad m$^{-2}$). At lower values of the Faraday rotation measure, the expected radio luminosity falls below the limit of detection threshold for present-day radio telescopes. These findings support the idea that the persistent radio sources observed so far are generated by a nebula in the FRB environment, and that  FRBs with low Faraday rotation measure may not show a persistent radio source because of a weaker magneto-ionic medium. This is generally consistent with models invoking a young magnetar as the central engine of the FRB, where the surrounding ionized nebula -  or the interacting shock in a binary system - powers the persistent radio source.}}

\frb\ is a repeating fast radio burst which underwent a reactivation during March 2021 \cite{ChimeAtel21,Lanman_2022}, allowing for an accurate localization with an uncertainty of a few milli-arcseconds through Very Long Baseline Interferometry (VLBI) observations \cite{2022ApJ...927L...3N}. Its precise localization (milli-arcsecond) and close distance (redshift $z =0.098$) with respect to other FRBs with an identified host galaxy make it an ideal target to study the physical conditions of its surroundings. Following the discovery of an extended radio emission associated with star formation in the FRB\,20201124A environment \cite{2021A&A...656L..15P}, we targeted the FRB region with deep, sub-arcsec angular resolution Very Large Array (VLA) observations at 15 and 22 GHz. A sub-arcsec resolution is crucial to disentangling a possible compact source from the previously detected diffuse (3--4 kpc) emission, allowing us to assess the presence of a persistent radio source (PRS) associated with the FRB.

\begin{figure}
    \centering
    \includegraphics[]{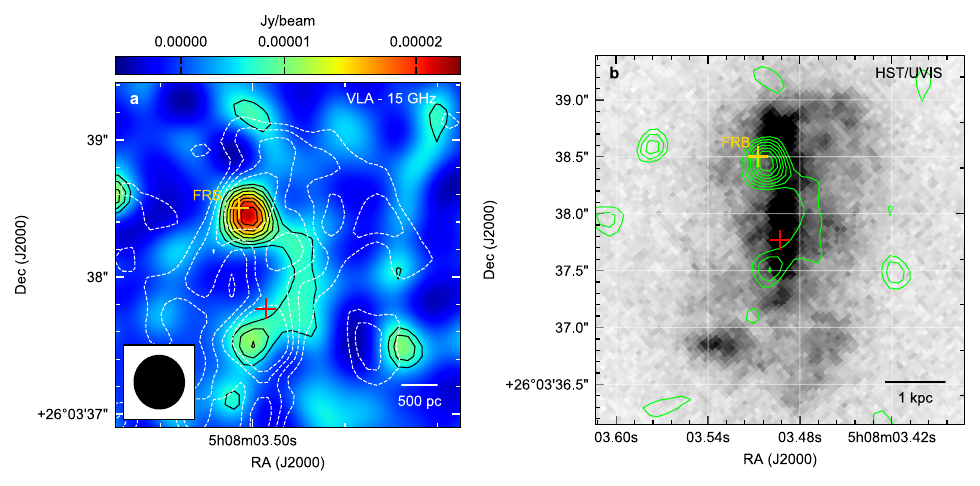}
    \caption{\textbf{Images of the host galaxy of \frb.} \textbf{a}, VLA image at 15 GHz from this work. Black contours indicate 3, 4, 5, 7, 9 $\times\sigma$ levels. The white contours are from the re-imaged 6 GHz VLA archival data\cite{2024ApJ...961...44D} at similar resolution, and indicate 3, 4, 5, 6 $\times\sigma$ levels. \textbf{b}, HST/UVIS image by \cite{2024ApJ...961...44D}, with contours reporting the 15 GHz VLA image from left panel. In both panels the location of the FRB as measured by the EVN \cite{2022ApJ...927L...3N} is reported as a yellow cross, while the center of the host galaxy as estimated from the HST/UVIS image \cite{2024ApJ...961...44D} ($F475X$ optical filter) is reported as a red cross. \label{fig:VLA+HST}}
\end{figure}

Observations at 15 GHz highlight a compact, unresolved emitting region, with a position consistent with the FRB location (see Fig. \ref{fig:VLA+HST}a). Its flux density outshines the diffuse component from the host galaxy at the same frequency. This bright, compact PRS falls in the South-West region of the extended emission previously discovered at higher frequencies \cite{2021A&A...656L..15P}. At 22 GHz, a compact component is still detected, although at a lower significance level of 4$\sigma$, at a position consistent with the FRB location and the peak of the 15 GHz component (see Extended Data Fig. \ref{fig:VLA+Dong}). 
The estimated projected physical size at 15 GHz is $\lesssim 700$ pc, and a much smaller size cannot be excluded. As also noted in \cite{2024ApJ...961...44D}, the flux densities measured up to 22 GHz imply that this compact source could not be detected in previous VLBI observations because it was too faint for the image noise level reached in those works -- $\sim$10 $\mu$Jy beam$^{-1}$ \cite{2021ATel14603....1M,2021A&A...656L..15P}, at the limit of capabilities for current VLBI arrays.

In order to derive the spectral shape of the source, we have re-imaged and re-analyzed the 6 GHz VLA archival data from ref.\cite{2024ApJ...961...44D} that have a similar resolution to our observations. We could disentangle the compact PRS from the diffuse star-formation emission, and estimate a flux density consistent with the one predicted by those authors on the basis of general image-quality considerations. Considering the flux densities collected from multi-frequency (6, 15, 22 GHz) VLA data, the PRS shows an inverted radio spectrum ($F_\nu\propto\nu^\alpha$, with $\alpha=1$).

%%%%%%%%%%%%%%%%%%%%%%%%%%%%%%%%%%%%%%%%%%%%%%%%%%%%%%

%%%%%%%%%%%%%%%%%%%%%%%%%%%%%%%%%%%%%%%%%%%%%%%%%%%%%%%%%%%%%%%%

%%%%%%%%%%%%%%%%%%%%%%%
% SFR 
%%%%%%%%%%%%%%%%%%%%%%%

\begin{figure}
    \centering
    \includegraphics{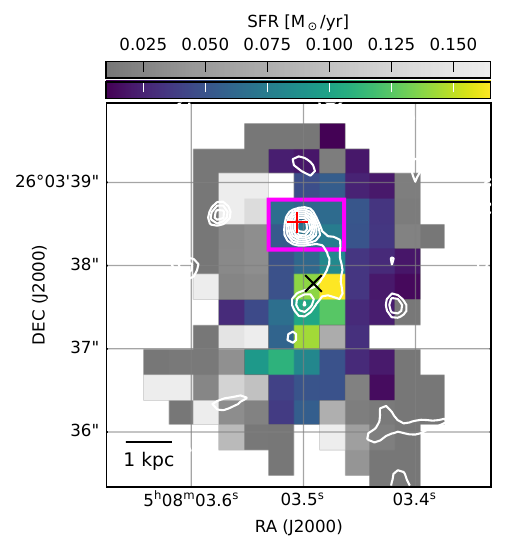}
    \caption{\textbf{Star formation rate map of the \frb\ host galaxy, as derived from GTC/MEGARA integral field spectroscopy.} This is estimated from the intrinsic H$\alpha$ luminosity, taking into account Galactic and intrinsic dust extinction using the Balmer decrement (H$\alpha$/H$\beta$). Pixels with an H$\alpha$ significance below a $3\sigma$ threshold are omitted. In cases where the H$\beta$ significance falls below a $2\sigma$ threshold -- so that no accurate correction for dust extinction is possible -- we give a lower limit of the SFR, and represent these pixels in grayscale. Note that the tick labels above the two colorbars represent lower limit SFRs and actual SFRs for grayscale and coloured pixels, respectively. The white contours represent the VLA image at 15 GHz, whereas the red and black crosses represent the FRB and host galaxy's centre, respectively.
    The magenta rectangle define the galactic region encompassing six adjacent MEGARA pixels surrounding the FRB and PRS zone.  \label{fig:megara_1}}
\end{figure}

%We investigated the possibility 
While the global, diffuse radio emission discussed by previous authors\cite{Fong21,2022MNRAS.513..982R,2021A&A...656L..15P,2024ApJ...961...44D} is linked to star-forming regions within the host galaxy (see Methods), it is difficult to associate the discovered compact radio source with the same physical process. First, the inverted spectrum is unlike the spectral shape observed in radio from star formation, usually steep ($\alpha<-0.5$) or at most flat ($-0.5<\alpha<-0.2$, see Methods). 
Furthermore, the radio luminosity of star forming regions is typically   three orders of magnitude lower than the luminosity of the compact radio source. Even the  rarest giant regions, with sizes $\lesssim 700$ pc, are ten times less luminous (see Methods).
Finally, even assuming a star formation rate capable of accounting for the radio luminosity, it would leave imprints at other wavelengths, that are not detected.
Our integral field unit observations with GTC/MEGARA (see Methods) confirms that the FRB falls at the Northern edge of the host galaxy's bar, as seen in previous HST images \cite{2024ApJ...961...44D}. The star formation rate (SFR) map, as estimated from the H$\alpha$ emission line (see Fig. \ref{fig:megara_1}), reveals that highest SFR is towards the host galaxy's center, and match the Optical emission as seen by HST (see Fig. \ref{fig:VLA+HST}b). Conversely, close to the FRB position, an SFR of $\sim0.4\,\rm{M}_\odot\,\rm{yr}^{-1}$ is measured.
%%%
Additionally, we performed mm-band observations with NOEMA centered on the FRB coordinates to independently estimate the SFR through dust grey-body emission. These resulted in two non-detections at 236.5 and 250 GHz, corresponding to $3\sigma$ upper limits of 130 $\mu$Jy beam$^{-1}$ and 160 $\mu$Jy beam$^{-1}$, respectively. The angular resolution of NOEMA images allowed to probe a physical scale of about 1.3$\times$0.7 kpc, comparable with the one probed by the VLA. We modelled the cold dust spectral energy distribution (SED) with a modified black body (MBB) function in order to derive upper limits for the dust mass and SFR (see details in Methods), finding a dust mass of $\rm{M}_{\rm dust}<7.9\times10^{6}\ {\rm \rm{M}_{\odot}}$ and a star formation rate of SFR$<2.2 \ \rm{M}_\odot\,yr^{-1}$ at the location of the 15 GHz PRS.
%%%
However, from our radio images we could estimate that, in order to account for the observed radio flux density, a ${\rm SFR}\gtrsim 3 \ \rm{M}_\odot\,yr^{-1}$ would be needed (see Methods). This number is not consistent with either the upper limit from obscured star formation derived from NOEMA observations (SFR$<2.2 \ \rm M_\odot\,yr^{-1}$), or the H$\alpha$-derived SFR at the location of the compact source as estimated from GTC/MEGARA spectroscopy ($\sim 0.4\,\rm{M}_\odot\,\rm{yr}^{-1}$), further supporting a different origin.

%%%%%%%%%%%%%%%%%%%%%%%%

%%%%%%%%%%%%%%%%%%%%%%%%%%%%%%%%%%%%%%%%%%%%%%%%%%%%%%%%%%%%%%%%

\begin{figure}
\centering
\includegraphics[width = 12cm, trim = 0 0 0 0, clip]{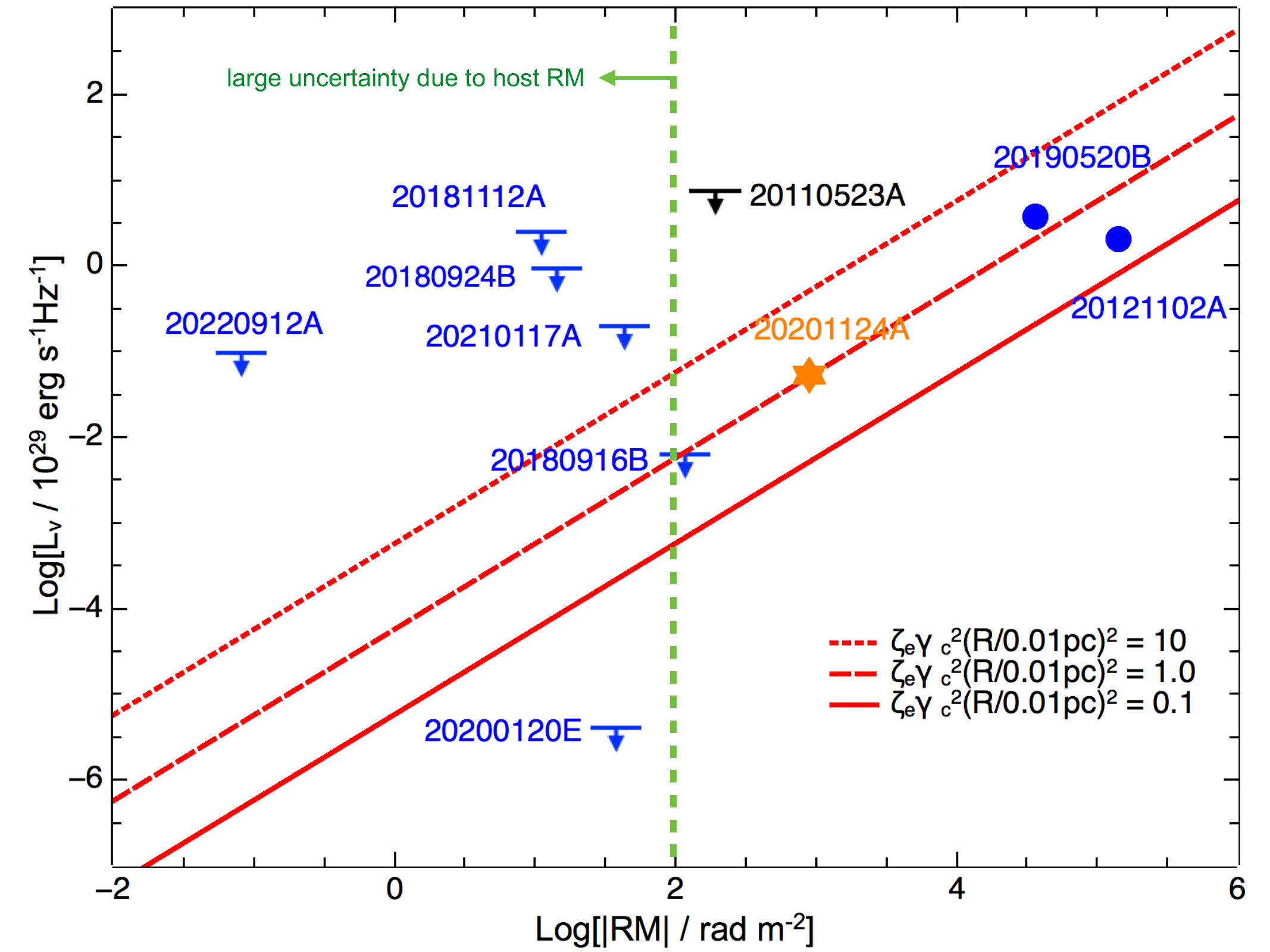}
\caption{\textbf{The proposed relation between the PRS specific radio luminosity and the FRB rotation measure \cite{yang20,yang22}.} The red dotted, dashed, and solid lines denote the predicted relation for $\zeta_e\gamma_{\rm c}^2(R/10^{-2}~{\rm pc})^2=10,1.0,0.1$, respectively. 
The orange star point indicates FRB 20201124A.
The black upper limit indicates the non-localized FRB 20110523A with an upper limit of the persistent emission and a measured value of RM, which gives one of the most conservative constraints for non-localized sources in literature. The blue circles denote the FRBs with measured persistent emission flux and RM (FRB 20121102A and FRB 20190520B). The blue upper limits correspond to the FRBs with precise (arcsecond) localizations.
Due to the large RM uncertainty of the host interstellar medium, the data of FRBs with $|{\rm RM}|\lesssim(10^2-10^3)~{\rm rad~m^{-2}}$ (on the left side of the green dashed line) might significantly deviate from the predicted relation.
\label{fig1}} 
\end{figure}

%%%%%%%%%%%%%%%%%%%%%%%%%%%%%%%%%%%%%%%%%%%%%%%%%%%%%%%%%%%%%%%%

%%%%%%%%%%%%%%%%%%%%%%%
% Nebula model 
%%%%%%%%%%%%%%%%%%%%%%%

Large Faraday rotation measure (RM) values, $|{\rm RM}|$, found for some FRBs imply a dense and magnetized environment in the vicinity of the sources \cite{yang20,yang22}. 
If relativistic electrons constitute a significant fraction of the medium, the magnetized environment near the FRB source could produce synchrotron radiation, powering a bright PRS \cite{Murase16,Metzger17,margalit18,yang20}.
There should be a simple relation between the luminosity of the PRS and the Faraday rotation measure, tracing the magneto-ionic medium in the FRB environment \cite{yang20,yang22}. We consider that the electron distribution in the PRS emission region has a thermal component and a non-thermal component, and the RM is mainly due to the thermal component.
As discussed in the Methods section, the foreseen relation is: \begin{equation}
L_{\nu}\propto(\zeta_e\gamma_{\rm c}^2R^2)\times|\rm RM|. 
\end{equation}
where $\zeta_e$ is the fraction of the electrons radiating synchrotron emission in the observed band among all electrons, $\gamma_{\rm c}$ is the typical Lorentz factor defined by $\gamma_{\rm c}^2=\int n_e(\gamma)d\gamma/\int[n_e(\gamma)/\gamma^2]d\gamma$, $n_e(\gamma)$ is the electron differential distribution with electron Lorentz factor $\gamma$, and $R$ is the radius of the plasma that contributes to the emission of the PRS and the RM.
 
In Figure \ref{fig1} we plot the relation between the specific luminosity (or upper limit) of the PRS against the RM of known FRBs. The red dotted, dashed, and solid lines correspond to the predicted relation for $\zeta_e\gamma_{\rm c}^2(R/10^{-2}~{\rm pc})^2=10,1,0.1$, respectively (see Methods).
The observed specific PRS luminosities of FRB\,20121102A, FRB\,20190520B, and FRB\,20201124A are close to the predicted value for $\zeta_e\gamma_{\rm c}^2(R/10^{-2}~{\rm pc})^2\sim(0.1-10)$. In particular, the detection of the PRS associated with FRB\,20201124A allows for the first time to test the predicted relation at lower radio luminosity, at the limits of present-day radio telescopes detection threshold, thus significantly broadening the probed parameter space.
The upper limit of FRB 20180916B is also close to the predicted relation. As a caveat, we should notice that for FRBs with ${\rm RM}\lesssim\text{a few }\times100~{\rm rad~m^{-2}}$, the RM from the Galactic component or the host component could dominate. This places FRB\,20201124A, with a $|\rm RM|\sim 900~{\rm rad~m^{-2}}$, at the limits of the explorable range.

These findings support a nebular origin for the three detected PRS associated with FRBs so far. 
While the possible nature of the FRB engine itself remains unclear, the  nebula model is generally consistent with the proposed scenarios involving a magnetar or a hyperaccreting X-ray binary as the central engine, which already predict an ionized nebula around it \cite{2018MNRAS.481.2407M,2022ApJ...937....5S}. Thanks to the increasing census of FRBs, and in particular repeating ones, deep radio observations of a larger sample of these objects could put further constraints on the nebula model, and on the physical conditions in the surroundings of FRBs in general. 

\

\bigskip
\subsection{Acknowledgments}
The research leading to these results has received funding from the European Union’s Horizon 2020 Programme under the AHEAD2020 project (grant agreement n. 871158)
B.O. gratefully acknowledges support from the McWilliams Postdoctoral Fellowship at Carnegie Mellon University.
Y.P.Y. is supported by the National Natural Science Foundation of China grant No. 12003028 and the National SKA Program of China (2022SKA0130100).
A.G. acknowledges financial support from the Severo Ochoa grant CEX2021-001131-S funded by MCIN/AEI/ 10.13039/501100011033 and from national project PGC2018-095049-B-C21 (MCIU/AEI/FEDER, UE).
The National Radio Astronomy Observatory is a facility of the
National Science Foundation operated under cooperative agreement by Associated Universities, Inc. 
This work is partly based on observations carried out under project number W22BS with the IRAM NOEMA Interferometer. IRAM is supported by INSU/CNRS (France), MPG (Germany) and IGN (Spain). 
Partly based on observations made with the Gran Telescopio Canarias (GTC), installed at the Spanish Observatorio del Roque de los Muchachos of the Instituto de Astrofísica de Canarias, on the island of La Palma. 
This work is partly based on data obtained with MEGARA/MIRADAS instrument, funded by European Regional Development Funds (ERDF), through Programa Operativo Canarias FEDER 2014--2020.
We wish to thank Armando Gil de Paz (Facultad Ciencias Físicas, Universidad Complutense de Madrid) for his valuable support to the MEGARA data analysis. 
This research made use of APLpy, an open-source plotting package for Python hosted at \url{http://aplpy.github.com}. This research made use of Astropy, a community-developed core Python package for Astronomy \cite{2013A&A...558A..33A}.

%%%%%%%%%%%%%%%%%%%%%%%%%%%%%%%%%%%

\subsection{Authors contribution}

G.B. led the VLA and NOEMA observational campaigns, conducted the VLA data calibration,
analysis and interpretation, and led the paper writing. L.P., Y.P.Y., B.Z., S.S. worked on the interpretation of the results. E.P., L.N., S.Q, A.M.N.G., A.R. conducted the GTC/MEGARA observations, data analysis, and interpretation. C.F., R.T. worked on the NOEMA data calibration and analysis. B.O. realized the host galaxy broad-band SED fitting. A.G. led the GTC/MEGARA proposal. R.P. contributed to the NOEMA proposal preparation.
All authors contributed to the discussion of the results presented and commented on the manuscript.

\subsection{Availability of data and materials}
All relevant data used for this work are publicly available at the repositories of each facility. In particular, raw and calibrated VLA data can be downloaded from the NRAO data archive (\url{https://data.nrao.edu/}), NOEMA raw data are available at the IRAM Science Data Archive (\url{https://iram-institute.org/science-portal/data-archive/}), and GTC raw data at The Gran Telescopio CANARIAS Public Archive (\url{https://gtc.sdc.cab.inta-csic.es/gtc/}).

%\subsection{Competing interests} The authors declare no competing interests.

%\section*{Supplementary information} 

%\subsection{Correspondence and requests for materials} 

%\section*{Peer review information}

%\section*{Reprints and permissions information}

%%%%%%%%%%%%%%%%%%%%%%%%%%%%%%%%%%%%%%%%%%%%%%%%%%%%%%%%%%%%%%%%%%%%%%%%%%%%%%%%%%%%%%%
%%%%%%%%%%%%%%%%%%%%%%%%%%%%%%%%%%%%%%%%%%%%%%%%%%%%%%%%%%%%%%%%%%%%%%%%%%%%%%%%%%%%%%%
\clearpage

\section*{METHODS}\label{sec11}

\textbf{VLA}
\label{appendix:VLA}
\\
Observations with the VLA were performed in B-configuration at 15 GHz (Ku-band) and 22 GHz (K-band), under DDT project 21B-330 [PI: Bruni]. The 22 GHz observations were divided into two $\sim2$-hours blocks, and then concatenated in order to achieve a similar RMS to the 15 GHz ones. A journal of observations can be found in Extended Data Tab. \ref{tab:vla_journal}. Data calibration was performed with the {\tt CASA}\footnote{\url{https://casa.nrao.edu/}, \cite{2022PASP..134k4501C}} VLA pipeline, and images deconvolution through the {\tt TCLEAN} task. The signal to noise ratio at 15 GHz allowed us to apply the {\textit{robust}} weighting scheme to improve the angular resolution, obtaining a final value for the FWHM similar to the 22 GHz one. The RMS was 2--3 $\mu$Jy/beam at both frequencies.

\noindent\textbf{NOEMA}
\label{app:NOEMA}
\\
Observations with the NOrthern Extended Millimeter Array (NOEMA) were carried-out through project W22BS [PI: Bruni] during March 2023 using the B array configuration with 12 antennas. A journal of observations can be found in Extended Data Tab. \ref{tab:vla_journal}. The phase center was set at RA 05:08:03.470, Dec 26:03:38.50. The Band 3 (1 mm) receiver was tuned at 244 GHz, so that the Lower Side Band (LSB) was centered at 236.5 GHz and the Upper Side Band (USB) at 250 GHz. 
Visibilities were calibrated using the {\tt CLID} pipeline within the {\tt GILDAS} software \footnote{\url{www.iram.fr/IRAMFR/GILDAS}}. The absolute flux scale accuracy is of the order 10\%. Imaging and deconvolution were performed using {\tt MAPPING} applying natural weighting. The final continuum maps reach a 1$\sigma$ RMS of $4.3 \times 10^{-5}$ Jy/beam in LSB and $5.3\times 10^{-5}$ Jy/beam in USB. The clean beams are $0\farcs70\times 0\farcs42$ and $0\farcs66\times 0\farcs39$  (PA = $15^\circ$) in LSB and USB, respectively.  

No clear detection was found in either LSB or USB data towards the FRB position, as derived from 15 GHz VLA data. We derived $3\sigma$ upper limits of 0.13 mJy and 0.15 mJy/beam in LSB and USB, respectively. 

%%%%%%%%%%%%%%%%%%%%%%%%%%%%%%%%%%%%%%%%%%%%%%%%%%%%%%%%%%%%%%%%%%%%%%%%%%%%

\noindent\textbf{GTC/MEGARA}
\label{sec:megara_appendix}
\\
The GTC/MEGARA instrument (\cite{Carrasco_2018}; \cite{Gil_de_Paz_2018}) collected a series of 6 observations on October 26, 2022, under project 99-GTC89/22B [PI: Gardini] using the Large Compact Bundle (LCB) Integral Field Unit (IFU) mode.
It provides a field of view (FoV) of $12\farcs5 \times 11\farcs3$ using 567 fibres in a hexagonal tessellation with a spaxel size of $0\farcs62$.
In order to cover the main optical emission lines, useful to derive the SFR spatial distribution within the galaxy, the observations were carried out with two different low resolution (LR, resolving power $\mathrm{R}\sim 6000$) volume-phase holographic (VPH) grating: VPH570-LR (LR-V), and VPH675-LR (LR-R), whose specifications can be found in in Extended Data Tab. \ref{tab:vla_journal}.
This produces a combined spectrum covering the spectral range from 4350 to 7288\;\AA.
%(see Figure~\ref{fig:MEGARA_total_spectrum}). 
This range, at the redshift of our target, includes spectral emission lines such as H$\alpha$, H$\beta$, $[\mathrm{O III}]$, $[\mathrm{N II}]$, needed for deriving the spatial distribution of the galaxy parameters like SFR, intrinsic extinction and velocity map. The measured seeing during the observations were in the range $0\farcs7$ -- $0\farcs9$.
We used the MEGARA data reduction pipeline (DRP) v0.12.0 \cite{sergio_pascual_2022_6043992} to process the data. 
We followed the same procedure as described in detail in \cite{Chamorro_Cazorla_2023}.
The final spectra where flux calibrated using the standard stars HR8634, available in the ESO spectro-photometric standards database (\cite{Oke_1990}).

\noindent\textbf{VLA detection of the compact PRS}
\\
%(See the observation log in Extended Data Tab. \ref{tab:vla_journal}).
We detected a compact component both at 15 and 22 GHz. Extended Data Fig. \ref{fig:VLA+Dong} shows the overplot of the image at 6-GHz that we have derived reanalyzing archival data from ref.\cite{2024ApJ...961...44D} with contours from our 15 and 22 GHz images at similar resolution. The emission comes from regions consistent with the EVN position at all frequencies. Indeed, a Gaussian fit results in a compact component at 15 and 22 GHz, with an integrated flux density of 20.0$\pm$3.5 $\mu$Jy and 30.0$\pm$9.7 $\mu$Jy, respectively. The peak position of the fitted Gaussian component is RA 05:08:03.50270$\pm$0.00097 s, Dec +26:03:38.444$\pm$0.014 arcsec at 15 GHz, while RA 05:08:03.4944$\pm$0.0052 s, Dec +26:03:38.423$\pm$0.080 arcsec at 22 GHz. Considering a standard astrometric accuracy of $\sim 10$\% of the FWHM, and summing that in quadrature with the Gaussian fit error, we get a final uncertainty of 0\farcs04$\times$0\farcs04 at 15 GHz, and 0\farcs08$\times$0\farcs09 at 22 GHz. The mentioned positional uncertainties imply that the detected compact component is consistent within 3$\sigma$ with the FRB location by ref.\cite{2022ApJ...927L...3N}. The measured flux density at 22 GHz accounts for $\sim 40$\% of the value reported at the same frequency, but at lower resolution, in ref.\cite{2021A&A...656L..15P}, indicating that the previously detected diffuse emission falls below the detection threshold per angular resolution element of the new image.

The 
%\cite{2024ApJ...961...44D} 
image at 6 GHz, and the 15 and 22 GHz ones presented here have a consistent angular resolution and RMS level, and can then be used to estimate the spectral index of the compact PRS. At 6 GHz, \cite{2024ApJ...961...44D} discuss how a possible PRS would be embedded in a diffuse emission region due to star formation. Those authors thus provide a conservative upper limit of 10 $\mu$Jy. We re-imaged the 6 GHz data from \cite{2024ApJ...961...44D} (program 22A-213, PI: Fong), and performed an additional analysis to refine the flux density and compactness estimate of the possible PRS. To account for the diffuse emission, we estimated the mean flux density in the regions contiguous with the FRB location, and outside a $\sim$1 beam area centered on the component peak at 6 GHz. In this way, we obtained an estimate of 8 $\mu$Jy/beam for the emission plateau due to star formation, consistent with the one discussed by \cite{2024ApJ...961...44D}. We then used this value as an input for the {\tt offset} parameter of the {\tt IMFIT} task, that allows to account for a zero-level offset when performing a Gaussian fit on the image. This procedure revealed a compact component at 6 GHz with an integrated flux density of $8.2\pm 3.8$ $\mu$Jy, and centered at RA 05:08:03.5007$\pm$0.0023 s, Dec +26:03:38.5440$\pm$0.0438 arcsec. As previously discussed, considering an astrometric accuracy of $\sim 10$\% of the FWHM we estimate a final uncertainty of 0\farcs05 $\times$ 0\farcs06, making again the peak position consistent with the FRB location within 3$\sigma$. We note that, despite the uncertainty on the component flux density is $\sim$45\% of the value, it is still at a 4$\sigma$ detection level with respect to the image RMS (2 $\mu$Jy/beam), and 8$\sigma$ considering its absolute peak of $\sim$16 $\mu$Jy given by the Gaussian+offset fitting. The estimated flux density is also consistent with the upper limit of 10 $\mu$Jy discussed by \cite{2024ApJ...961...44D}.

Considering the estimated flux densities at 6, 15, and 22 GHz, we obtain the following spectral indices ($S\propto\nu^\alpha)$ for the PRS: $\alpha_{6}^{15}=0.97\pm0.54$, $\alpha_{15}^{22}=1.06\pm 0.96$, and an overall value $\alpha_{6}^{22}=1.00\pm0.43$, indicating an inverted spectrum.

%%%%%%%%%%%%%%%%%%%%%%%%%%%%%%%%

\noindent\textbf{SFR estimates from NOEMA observations}
\\
%(See the observation log in Extended Data Tab. \ref{tab:vla_journal}).
In order to derive upper limits for the dust mass and SFR at the location of the PRS based on the NOEMA fluxes, we modelled the SED of the dust emission with a modified black-body (MBB) function given by
 \begin{equation}
 \label{eq:sed}
     S_{\nu_{\rm obs}}^{\rm obs} = S_{\nu/(1+z)}^{\rm obs} = \dfrac{\Omega}{(1+z)^3}[B_{\nu}(T_{\rm dust}(z))-B_{\nu}(T_{\rm CMB}(z))](1-e^{-\tau_{\nu}}), 
 \end{equation}
\noindent where $B_{\nu}$ is the black body function that depends on the temperature and rest frequency ($\nu$), $\Omega = (1+z)^4A_{\rm gal}D_{\rm L}^{-2}$ is the solid angle with $A_{\rm gal}$ , and $D_{\rm L}$ is the surface area and luminosity distance of the galaxy, respectively. The dust optical depth is
\begin{equation}
    \tau_{\nu}=\dfrac{M_{\rm dust}}{A_{\rm galaxy}}k_0\biggl(\dfrac{\nu}{250\ \rm GHz}\biggr)^{\beta},
\end{equation}
\noindent with $\beta$ the emissivity index and $k_0 = 0.45\  \rm cm^{2}\ g^{-1}$ the mass absorption coefficient \cite{beelen+2006}. The solid angle is estimated using the size of the NOEMA beam, that is similar for both observations, $0\farcs7\times 0\farcs4$. The effect of the CMB on the dust temperature is given by
\begin{equation}
    T_{\rm dust}(z)=((T_{\rm dust})^{4+\beta}+T_0^{4+\beta}[(1+z)^{4+\beta}-1])^{\frac{1}{4+\beta}},
\end{equation}
\noindent with $T_0 = 2.73$ K.
We also considered the contribution of the CMB emission given by $B_{\nu}(T_{\rm CMB}(z)=T_0(1+z))$ \cite{dacunha2013}, even if at this redshift it is almost negligible.

%%%%%%%%%%%%%%%%%%%%%%%%%%%%%%%%
%

Therefore, this model has three free parameters: $T_{\rm dust}$, $M_{\rm dust}$, and $\beta$. Since we were able to determine only upper limits in the two NOEMA bands, we evaluated possible combination of $T_{\rm dust}$ and $\beta$ that can reproduce the limits at 236 GHz and 250 GHz, leaving $M_{\rm dust}$ as a free parameter. Extended Data Fig. \ref{fig:sed-submm}a presents 4 models built from a reasonable combinations of dust temperature and emissivity: $T_{\rm dust}=20, 30$ K and $\beta=1.5,2.0$, that are typical values found in local star-forming galaxies \cite{schreiber2018,lamperti2019}. We explored the one dimensional parameter space using a Markov chain Monte Carlo (MCMC) algorithm implemented in the \texttt{EMCEE} package \cite{foreman2013}, and we assumed a uniform prior for the dust mass, $10^{4} \ {\rm M_{\odot}}<M_{\rm dust}<10^{9}\ {\rm M_{\odot}}$. We obtained $M_{\rm dust}/{10^6 \rm M_{\odot}} = 7.9, 7.8, 4.7, 4.5$ with $[T_{\rm dust} (\rm K), \beta]=[20, 1.5], [20,2.0], [30,1.5], [30,2.0]$, from a MCMC with 25 chains, 1500 trials and a burn-in phase of $\sim 50$ for each model.

The SFR is estimated from the total infrared luminosity (TIR) as ${\rm SFR} [{\rm M_\odot ~yr^{-1}}] = 10^{-10} \ L_{\rm TIR} \rm [L_\odot]$, assuming Chabrier IMF \cite{Chabrier2003}. To compute the TIR luminosity we integrated the MBB from $8$ to $1000\ \mu$m rest-frame, and therefore we obtained SFR = 0.13, 0.34, 0.69, and 2.24 $\rm M_\odot ~yr^{-1}$ for $[T_{\rm dust} (\rm K), \beta]=[20, 1.5], [20,2.0], [30,1.5], [30,2.0]$, respectively. Overall we obtained that $M_{\rm dust}<7.9\times10^{6}\ {\rm M_{\odot}}$ and SFR$<2.2 \;\rm M_\odot ~yr^{-1}$.

In order to constrain the emission from the host galaxy (see also ref.\cite{2024ApJ...961...44D}), we re-computed the $3 \sigma$ upper limits at 236 GHz and 250 GHz considering the region with significance $>3 \sigma$ in the 6 GHz VLA map, which corresponds to a physical region of $\sim 2.85$ kpc$^2$. The $3 \sigma$ upper limits are then $0.49$ mJy and $0.55$ mJy at 236 GHz and 250 GHz, respectively. Panel \textbf{b} of Extended Data Fig. \ref{fig:sed-submm} shows the best-fitting curves to the new upper limits for $[T_{\rm dust} (\rm K), \beta]=[20, 1.5], [20,2.0], [30,1.5], [30,2.0]$. We obtained $1.6 <M_{\rm dust}/{10^7 \rm M_{\odot}}< 2.9$, implying ${\rm SFR} < 8.8\ \rm M_\odot ~yr^{-1}$. 

%%%%%%%%%%%%%%%%%%%%%%%%%%%%%%%%%%%%%%%%%%%%%%%%%%%%%%%%%%%%%%%%%%%%%%%%%%%%
%%%%%%%%%%%%%%%%%%%%%%%%%%%%%%%%%%%%%%%%%%%%%%%%%%%%%%%%%%%%%%%%%%%%%%%%%%%%

\noindent\textbf{SFR estimates from GTC/MEGARA observations}
\\
%(See the observation log in Extended Data Tab. \ref{tab:vla_journal}).
Since the emission lines observed in the FRB region and the entire galaxy spectra are produced by UV radiation from the stellar photosphere (see Supplementary Information), we can thus estimate SFRs directly from the H$\alpha$ intrinsic luminosity. 
First, we corrected H$\alpha$ for intrinsic dust attenuation based on the Balmer decrement.
Extended Data Fig. \ref{fig:megara_halpha+ebv+lines} shows the emission line maps together with the E(B$-$V) map of the galaxy obtained from the Balmer decrement. The measured line fluxes are reported in Extended Data Tab. \ref{tab:megara_stacked}.
To perform a proper correction for dust extinction, in pixels with S/N(H$\beta$) $\geq 2$, the colour excess E(B$-$V) is derived adopting the ref.\cite{Calzetti2000} attenuation law and assuming the Case B recombination and a Balmer decrement H$\alpha$/H$\beta=2.86$ (typical of H II regions with electron temperatures Te = $10^4$ K and electron density n$_e$ between $10^2$ -- $10^4$ cm$^{-3}$ \cite{Osterbrock2006}).
We assigned E(B$-$V) = 0 to pixels with negative E(B$-$V) values between about $-0.05$ and $\sim 0$ (i.e. pixels showing $\sim 2.7 \leq\; $H$\alpha$/H$\beta < 2.86$). 
Moreover, we also assigned an E(B$-$V) lower limit to pixels with S/N(H$\beta$) $<2$, using the upper limit in H$\beta$ we defined earlier in this section. 
The dust-corrected fluxes are converted to luminosity.
Then, we derived the SFR using the relation by ref.\cite{Kennicutt1998} and the initial mass function (IMF) by ref.\cite{Chabrier2003}.
We have already shown and discussed the resulting SFR map of the galaxy in Fig. \ref{fig:megara_1}.
To summarize, we report that in the stacked spectrum of the FRB region (Fig. SPECTRA), we measured an SFR of $0.41^{+0.02}_{-0.01}$ M$_\odot$/yr, while the entire galaxy has an SFR of $4.11^{+0.59}_{-0.53}$ M$_\odot$/yr.

%%%%%%%%%%%%%%%%%%%%%%%%%%%%%%%%%%%%%%%%%%%%%%%%%%%%%%%%%%%%%%%%%%%%%%%%%%%%

\noindent\textbf{Expected radio emission from a star forming region}
\label{sec:sfregion}
\\
Radio emission from SF regions involves typically  two components: a steep component ($\approx \nu^{-0.7}$) produced by synchrotron emission from electrons accelerated in SN explosion, with a lifetime of the order of 100 Myr \cite{Condon92}, that dominates the emission at low frequencies (below 6 GHz); a thermal, flat ($\approx \nu^{-0.5\leq\alpha\leq-0.2}$) component  by Bremmstrahlung, that is a direct measure of the current production of ionized photons by young and hot stars with lifetime $< 10$ Myr \cite{1981A&A....94...29K,1982A&A...116..164G,2017ApJ...836..185T}.  On the other hand, the inverted spectrum observed in the compact radio source is not consistent with either SF-driven radio components. In Fig. \ref{fig:radio-SED} we compare the radio spectral slopes for the PRS presented in this work, the host galaxy's nucleus, and the global host galaxy's emission. The nucleus's flux densities were collected using the data from \cite{2024ApJ...961...44D} (6 GHz) and this work (15, 22 GHz), while for the total radio emission of the host galaxy we used the lower resolution VLA observations from our previous work\cite{2021A&A...656L..15P}, capable of recovering the emission from extended regions. The slopes clearly indicate a different origin for the emission of the PRS, not consistent with the global steep spectrum of the host galaxy ($\alpha=-0.7$, as also discussed in Supplementary Information), or the nuclear one ($\alpha=-0.5$).

Furthermore, the radio luminosity of star forming regions in either our and external galaxies is about three orders of magnitude lower than the luminosity of the compact radio source. Even the  brightest giant regions, with sizes $\lesssim 700$ pc, are ten times less luminous\cite{1984ApJ...287..116K,2004MNRAS.355..899C,2014ApJS..212....1A,2011ApJS..194...32A}.
Finally, even in the implausible assumption of a star forming origin, the level of SFR required to account for the observed radio emission would violate the limits derived from mm and H$_{\alpha}$ measurements. The SFR needed to produce the radio thermal component is \cite{murphy11}: 
\begin{equation}
\frac{{\rm SFR}^T}{\Msolar yr^{-1}}= 4.6\ 10^{-28} \frac{T_e}{10^4 K} \nu_{GHz}^{0.1} \frac{L(\nu_{GHz})}{erg\; s^{-1}\, Hz^{-1}}
\end{equation}
%i.e. the lack of a  steep component produced by sycnhrotron emission, would require by a very young SF region, less than a few million years.  
The observed $L_{15 \rm{GHz}}= 4.9\times 10^{27}\,$erg\;s$^{-1}\,$Hz$^{-1}$ would then  imply
${\rm SFR}^T\gtrsim 3 \Msolar$~yr$^{-1}$ to account for  the radio emission.  This number is not consistent with either the upper limit from obscured star formation derived from NOEMA or from the H$\alpha$ measurement at the location of the compact source. 
In principle, the limit from the H$\alpha$ measurement could be alleviated by positing an  obscured star formation, but this looks rather unlikely for the following reasons. The fraction of the obscured star formation  for  this specific  galaxy, derived by comparing  the SFRs derived from the Ha luminosity and  the radio luminosity from the whole galaxy is about 60\%,  while at the FRB position an obscured fraction of about 90\% would be required to account for the radio luminosity. There is no evidence of a stronger obscuration towards the FRB, on the contrary  the extinction in that direction  is  a factor of two lower than towards the center of the galaxy.    

%
%%%%%%%%%%%%%%%%%%%%%%%%%%%%%%%%%%%%%%%%%%%%%%%%%%%%%%%%%%%%%%%%
%

\noindent\textbf{Nebular emission model}\label{sec3} 
\\
We consider that the electron Lorentz factor $\gamma$ has a differential distribution  $n_e(\gamma)$. The total electron number density is then $n_{e,0}=\int n_e(\gamma)d\gamma$. The RM contribution from relativistic electrons is suppressed by a factor of $\gamma^2$ due to the relativistic mass $m_e\rightarrow\gamma m_e$\cite{qua00}, thus, the RM contributed by the electrons with the distribution $n_e(\gamma)$ could be approximately written as
\be
{\rm RM}
\simeq\frac{e^3}{2\pi m_e^2c^4}B_\parallel\Delta R\int \frac{n_e(\gamma)}{\gamma^2}d\gamma
=\frac{e^3}{2\pi m_e^2c^4}\frac{n_{e,0} B_\parallel}{\gamma_{\rm c}^2} \Delta R,\label{RM}
\ee
where $B_\parallel$ is the mean parallel component of the magnetic field along the line of sight, $\Delta R$ is the thickness of the nebula, and the Lorentz factor $\gamma_{\rm c}$ is defined by
\be
\gamma_{\rm c}^2\equiv\frac{\int n_e(\gamma)d\gamma}{\int [n_e(\gamma)/\gamma^2]d\gamma}.
%=\frac{n_{e,0}}{\int [n_e(\gamma)/\gamma^2]d\gamma}
\ee
The Lorentz factor of the electrons emitting synchrotron radiation in the observation frequency $\nu$ is required to be
\be
\gamma_{\rm obs}\sim\left(\frac{2\pi m_ec\nu}{eB}\right)^{1/2}\simeq600\left(\frac{\nu}{1~{\rm GHz}}\right)^{1/2}\left(\frac{B}{1~{\rm mG}}\right)^{-1/2}. \label{gammaghz}
\ee
We define $\zeta_e$ as the number fraction of the electrons radiating synchrotron emission in the observed band in all electrons. It can be approximately given by
\be
\zeta_e\sim\frac{\gamma_{\rm obs}n_e(\gamma_{\rm obs})}{n_{e,0}}.
\ee
For example, if the electron distribution has a thermal component satisfying three-dimensional Maxwell distribution in the low-energy regime and a power-law component in the high-energy regime, one would approximately have\cite{yang22} $\zeta_e\sim(\gamma_{\rm obs}/\gamma_{\rm c})^{1-p}$, where $p$ is the distribution index of the high-energy component.
Assuming that the magnetic field is large-scale (leading to the parallel mean magnetic field of the order of the total magnetic field)
the specific luminosity of synchrotron radiation could be written as \cite{yang20,yang22}
\begin{align}
&L_{\nu}= \frac{64\pi^3}{27}\zeta_e\gamma_{\rm c}^2m_ec^2R^2\left|{\rm RM}\right|\simeq5.7\times10^{28}~{\rm erg~s^{-1}~Hz^{-1}} \nonumber \\
&\times\zeta_e\gamma_{\rm c}^2 \left(\frac{\left|{\rm RM}\right|}{10^4~{\rm rad~m^{-2}}}\right)\left(\frac{R}{10^{-2}~{\rm pc}}\right)^2 \label{lum} 
\end{align}
where $R$ is the radius of the plasma screen that contributes to the PRS and the RM. 

In Extended Data Tab. \ref{tab:tab1} we list 16 FRBs with the measured RM and the measured values (or upper limits) of the flux densities of the PRSs. Only FRB 20121102A, FRB 20190520B, and FRB 20201124A have measured values of both. 
Some sources have precise localizations, so the redshifts in the top row of Extended Data Tab. \ref{tab:tab1} are the directly measured values from their host galaxies. 
For other FRBs, due to the lack of a precise localization, we estimate their redshifts and luminosity distances via the extragalactic DM as performed by \cite{yang20}.
Notice that for redshift inferred from the excess DM, the corresponding redshift error is attributed to the IGM density fluctuation \cite{mcq14,yang20}. 

Based on the data in Extended Data Tab. \ref{tab:tab1}, we plot the relation between the specific luminosity of the PRS and the RM of FRBs in Figure \ref{fig1} where we only list some sources with strong constraints.
The red dotted, dashed, and solid lines correspond to the predicted relation for $\zeta_e\gamma_{\rm c}^2(R/10^{-2}~{\rm pc})^2=10,1.0,0.1$, respectively.
It is interesting that the observed specific luminosities of FRB 20121102A, FRB 20190520B, and FRB 20201124A are close to the predicted value for $\zeta_e\gamma_{\rm c}^2(R/10^{-2}~{\rm pc})^2\sim(0.1-10)$. 
The upper limit of FRB 20180916B is close to the predicted relation. However, we should notice that for the FRB with ${\rm RM}\lesssim\text{a few }\times100~{\rm rad~m^{-2}}$, the RM from the Galactic component or the host component would dominate.

%%%
\noindent\textbf{Considerations on the observed radio spectrum}
\\
As pointed out above, the PRS spectrum of FRB 20201124A satisfies $F_\nu\propto\nu^{1}$, which is much harder than those of FRB 20121102A and FRB 20190520B (see Extended Data Fig. \ref{fig:radio-SED} in  for a comparison). 
If the observed spectrum of the PRS is produced by synchrotron radiation, a spectral index larger than $1/3$ would imply that  synchrotron absorption might be dominant at the observed bands $(6-22)\;{\rm GHz}$. Such a high absorption frequency would lead to observed FRBs which absorbed by the PRS, which is inconsistent with the observation of FRB 20201124A\footnote{For example, for a certain FRB repeater, the absorption process would lead to the spectrum of a radio burst suppressed at a frequency lower than the absorption frequency. This implies that the radio bursts emitted during a limited time would have the same low-frequency cutoff. However, such a feature has not been found so far even for FRB 20201124A.}. 
On the other hand, we notice that the spectral index of the PRS is also consistent with $1/3$ within $1.5\sigma$. 
%Meanwhile, the upper limits of NOEMA constrain the spectral index to $\alpha\lesssim0.6$ at $(6-240)~{\rm GHz}$, which is less than that at $(6-22)~{\rm GHz}$. 
In this case, to explain the spectral shape of the PRS, the typical frequency $\nu_m$ that corresponds to the minimum Lorentz factor of the accelerated non-thermal electrons required to be
\begin{align}
\nu_m\simeq\frac{\gamma_m^2eB}{2\pi m_ec}\gtrsim(22-240)~{\rm GHz},
\end{align}
where $B$ is the magnetic field strength at the emission region. Therefore, we obtain the following constraints:
\begin{align}
\left(\frac{\gamma_m}{10^3}\right)^2\left(\frac{B}{10~{\rm mG}}\right)\gtrsim(0.8-8.6).\label{gammaB}
\end{align}
FRB 20201124A has an RM fluctuation of $\delta{\rm RM}\sim200~{\rm rad~m^{-2}}$ on a timescale of 10 days \cite{xu22}. Its DM fluctuation is not well measured but constrained to be 
$\delta{\rm DM}\lesssim3~{\rm pc~cm^{-3}}$ on a similar timescale. Thus, one can infer $B_\parallel > 0.1~{\rm mG}$, which is consistent with the above result.
The compact PRS associated with FRB 20201124A might be a magnetized, shocked nebula or an interacting shock in a binary system due to the following reasons: 
1) The observed non-thermal emission of the PRS is more likely produced by synchrotron radiation of the accelerated relativistic electrons, and shocks are natural locations to accelerate these particles. 
2) The typical magnetic field strength of $B\sim10~{\rm mG}$ is much higher than $B\sim1~{\rm \mu G}$ of the interstellar medium \cite{Draine11}. The observations of supernova remnants show that their magnetic field strength is typically $B\sim{\rm a~few~\mu G}$ to a few mG \cite{Reynolds12}. In a binary system, the magnetic field strength in the stellar wind could also be much larger than that in the interstellar medium.
3) There are two main possible scenarios involving a shock region in an astrophysical environment: (1) the interaction between an ejecta and interstellar medium (e.g., supernova remnants, afterglows of gamma-ray bursts) or between an electron-positron pair wind and an ejecta (e.g., pulsar wind nebulae); (2) an interacting shock between a pair wind and a companion wind in a binary system. The former corresponds to a magnetized shocked nebula and the latter requires that the FRB source is in a binary system. 
4) Other radically different combinations of $\gamma_mB$ seem not easy to implement in real astrophysical environments. For example, a much lower $\gamma_m\sim10$ leads to $B\sim100~{\rm G}$ according to Eq.(\ref{gammaB}). Such a high magnetic field might be in the magnetosphere or the pair wind of a neutron star as the FRB source, but the electrons' Lorentz factor in these regions should be much higher than $\gamma_m\sim10$. On the other hand, a much lower magnetic field $B\sim1~{\rm \mu G}$ requires $\gamma_m\sim10^5$, which means that there are many extremely relativistic electrons in the interstellar medium. However, if these extremely relativistic electrons exist, they should emit and become cool in the near-source environment with a stronger magnetic field.

At last, if repeating FRBs are indeed powered by a young flaring magnetar embedded within a supernova remnant, an expanding magnetized electron–ion nebula created by the outbursts could naturally contribute to the PRS and the large burst's RM \cite{margalit18}.
Since such a time-dependent model predicts a secular power-law
decay of the RM and PRS luminosity, considering that the source number should be proportional to the source age, the RMs from a sample of FRB nebulae are predicted to be distributed logarithmically, which could be tested in the future.

%%%%%%%%%%%%%%%%%%%%%%%%%%%%%%%%%%%%%%%%%%%%%%%%%%%%%%%%%%%%%%%%%%%%%%%%%%%%

\noindent\textbf{Considerations on a possible background source}
\\
In the GTC/MEGARA spectrum, we found a detection of a possible emission line in correspondence with the FRB location. If identified as [OII], the emitting source would be at redshift  $z=0.549$. We thus tested the scenario of a background source potentially responsible for the compact radio component, that we instead interpreted as a PRS linked to the FRB. 

Deep radio surveys\cite{2000ApJ...533..611R} show that the radio sky at a flux level of $\approx 10\;\mu$Jy exhibit about $1.4\times10^{-3}$ sources arcsec$^{-2}$. The vast majority has a steep spectrum,  unlike the one displayed by the source we observe here. Taking into account that only $\lesssim 7\%$ has a flat spectrum, we derive that the probability of one of them falling within the beam size at the position of the FRB is $\approx 10^{-5}$. This already results in a very low chance that the radio emission is due to a serendipitous background source. However, in the following we discuss the different classes of sources that could produce the compact radio component studied in this work, and why we can confidently rule them out.

Possible extragalactic sources producing radio emission are blazars, radio galaxies, radio quiet AGN, or a star-forming galaxy. At the mentioned redshift ($z=0.549$), the radio luminosity as estimated from our VLA measurement would be $L_R\sim10^{39}$ erg s$^{-1}$, thus indicating a radio-quiet regime. This already rules out a possible background blazar or radio galaxy, although it is still consistent with the faintest (radio quiet) Seyfert galaxies observed to date in the radio band ($10^{35}<L_R<10^{39}$ erg s$^{-1}$, ref.\cite{2019MNRAS.485.3185C}). A radio quiet AGN could show a radio component due to a low-power jet, a nuclear wind, or coronal emission \cite{2019NatAs...3..387P}. In the first case, while the physical scale subtended at $z=0.549$ by the compact radio component ($\sim 2$ kpc) would be enough to accommodate the presence of a jet, the expected radio spectrum of its extended regions would be steep ($\alpha<-0.5$). The inverted radio spectrum we found would thus require a collimated jet with an orientation within a few degrees from the line of sight. This would in turn result in a dominant, Doppler-boosted jet emission visible up to the optical band. This feature has been commonly observed for powerful jets at different scales, from radio-loud AGN to X-ray binaries, but not yet in radio-quiet AGN where the flat/inverted radio spectrum is attributed to a coronal emission mechanism (ref.\cite{2018MNRAS.478..399B,2023MNRAS.525..164C}, see also discussion below). Moreover, the NOEMA non-detection in the mm band put a strong constraint on that: a simple power-law extrapolation of the radio spectrum up to 250 GHz results in an expected flux density $>$300 $\mu$Jy, well above the NOEMA upper limit ($<$160 $\mu$Jy). %Alternatively, the radio spectrum should flatten at frequencies $>$30 GHz, leading to a mm-band flux density lower than the NOEMA upper-limit, implying that the inverted spectrum observed in the GHz range is due to synchrotron self-absorption. While the latter can be present in the most compact regions of the inner jet and radio cores, a spectrum showing a peak at frequencies $>$50 GHz would require a sub-pc physical size, comparable with the one of the corona \cite{2018MNRAS.478..399B}, thus not consistent with a jet (typically larger than several pc).
In the second case (nuclear wind), a steep radio spectrum is expected, thus not consistent with the inverted one seen in our VLA measurements. In the third case (AGN coronal emission), a ratio between radio and X-ray luminosity $L_R/L_X\sim10^{-5}$ is expected \cite{2008MNRAS.390..847L}, while previous Chandra X-ray measurements in a circular region with a 1" radius centered on the FRB location \cite{2021A&A...656L..15P} constrain this ratio to $\sim10^{-3}$. The X-ray luminosity is thus much lower than the one expected from a corona able to produce the radio emission measured with the VLA. 
Finally, a possible radio emission from a background star-forming galaxy is excluded by the expected relation between [OII] and radio emission in SFR \cite{Kennicutt1998}, foreseeing a radio luminosity  $\sim$10 times  lower than the one measured here.      

\clearpage
\setcounter{figure}{0}   
\setcounter{table}{0} 
%\captionsetup[table]{name=Extended Data Tab.}
%\captionsetup[figure]{name=Extended Data Fig.}

\section*{EXTENDED DATA}\label{sec_ED}

\begin{EDfigure}
    \centering
    \includegraphics[]{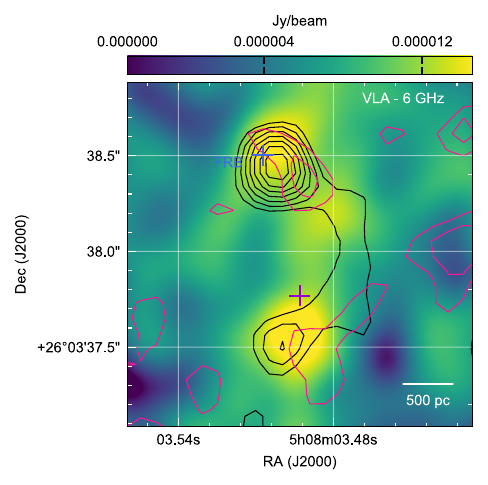}
    \caption{\textbf{Multi-frequency radio image of the host galaxy of \frb}. Overplot of the 6 GHz VLA image from \cite{2024ApJ...961...44D} (in colors) with contours from the 15 (black) and 22 (red) GHz VLA images from this work. The purple cross indicated the host galaxy center, while the blue cross is the FRB position.\label{fig:VLA+Dong}}
\end{EDfigure}

\clearpage

%%%%%%%%%%%%%%%%%%%%%%%%%%%%%%%%

\begin{EDfigure}
    \centering
    \includegraphics[]{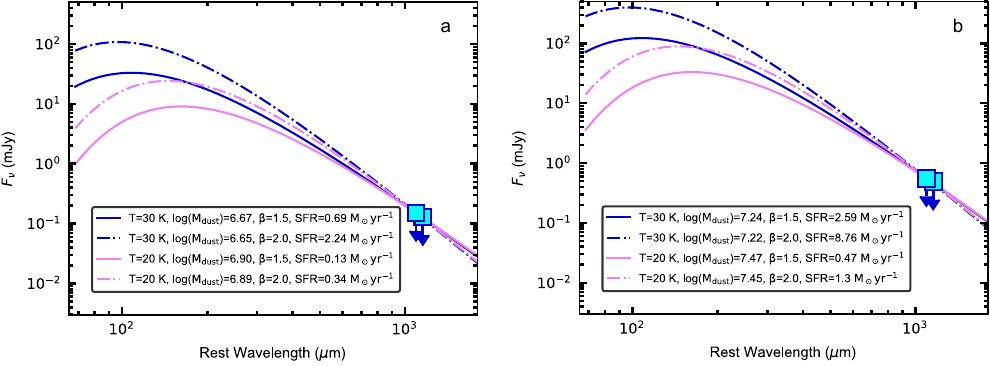}
    \caption{\textbf{SEDs of the cold dust emission of FRB 20201124A}. Estimated values are based on the NOEMA upper limits at 236 GHz and 250 GHz (cyan squares). \textbf{a}, SED computed in one beam towards the FRB position as seen in the VLA 15 GHz map. The best fitting curves with fixed $[T_{\rm dust} (\rm K), \beta]=[20, 1.5], [20,2.0], [30,1.5], [30,2.0]$ are plotted as solid violet, dashed violet, solid blue and dashed blue lines. \textbf{b}, SED of the cold dust emission of FRB 20201124A based on the NOEMA upper limits at 236 GHz and 250 GHz (cyan squares) computed for the region with $> 3 \sigma$ in the 6 GHz map. The best fitting curves with fixed $[T_{\rm dust} (\rm K), \beta]=[20, 1.5], [20,2.0], [30,1.5], [30,2.0]$ are plotted as solid violet, dashed violet, solid blue and dashed blue lines. \label{fig:sed-submm}} %\label{fig:sed-submm-bigarea}}
\end{EDfigure}

\clearpage

%\begin{figure}
%    \centering
%    \includegraphics[width=8cm]{sed-FRB20201124A-piro-nature.pdf}
%    \caption{SED of the cold dust emission of FRB 20201124A based on the NOEMA upper limits at 236 GHz and 250 GHz (cyan squares) computed for the region with $> 3 \sigma$ in the 6 GHz map. The best fitting curves with fixed $[T_{\rm dust} (\rm K), \beta]=[20, 1.5], [20,2.0], [30,1.5], [30,2.0]$ are plotted as solid violet, dashed violet, solid blue and dashed blue lines.\label{fig:sed-submm-bigarea}}
%\end{figure}

\clearpage
%%%%%%%%%%%%%%%%%%%%%%%%%%%%%%%%

\begin{EDfigure}
    \centering
    \includegraphics{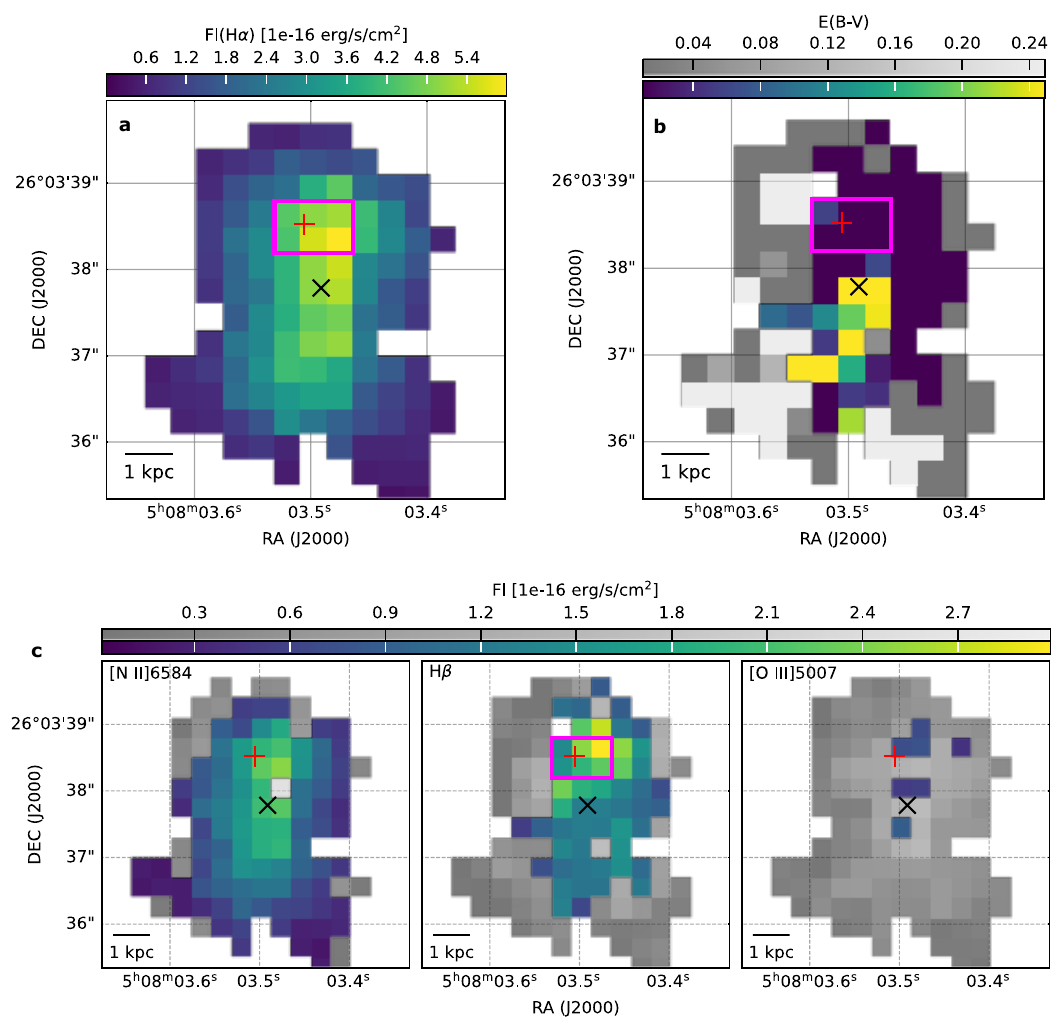}
    \caption{\textbf{GTC/MEGARA maps}.
    \textbf{a}, Galactic extinction corrected H$\alpha$ emission line map.
    \textbf{b}, Map of the intrinsic E(B$-$V) of the galaxy. We estimated the intrinsic E(B$-$V) taking into account Galactic dust extinction and using the Balmer decrement (H$\alpha$/H$\beta$). In cases where the H$\beta$ significance falls below a $2\sigma$ threshold no accurate correction for dust extinction is possible. For these pixels we give a lower limit for the intrinsic E(B$-$V) and are shown in grayscale. Note that the tick labels above the two colorbars represent lower limits and actual E(B$-$V) values for grayscale and coloured pixels, respectively.
    \textbf{c}, Maps of the emission lines [N II] 6584, H$\beta$, and [O III] 5007.
    In cases where the line significance falls below a $2\sigma$ we give a line upper limit, and represent these pixels in grey-scale.
    The tick labels above the two colorbars represent upper flux limits and actual flux values for grayscale and coloured pixels, respectively.
    The red and black crosses represent the FRB and galactic centre, respectively.
    %, as in Figure \protect\ref{fig:VLA+HST}.
    The magenta rectangle defines the galactic region encompassing six adjacent MEGARA pixels surrounding the FRB and PRS zone.
    For all the maps, pixels with an H$\alpha$ significance below a $3\sigma$ threshold are omitted.
    %\label{fig:megara_ebv}
    \label{fig:megara_halpha+ebv+lines}}
\end{EDfigure}

\clearpage

\begin{EDfigure}
    \centering
    %\hspace*{0.2cm}
    %\includegraphics[width=16cm]{spectra_FRB_stacking_v1.pdf}
    %\vspace*{-0.2cm}
    %\hspace*{0.2cm}
    \includegraphics{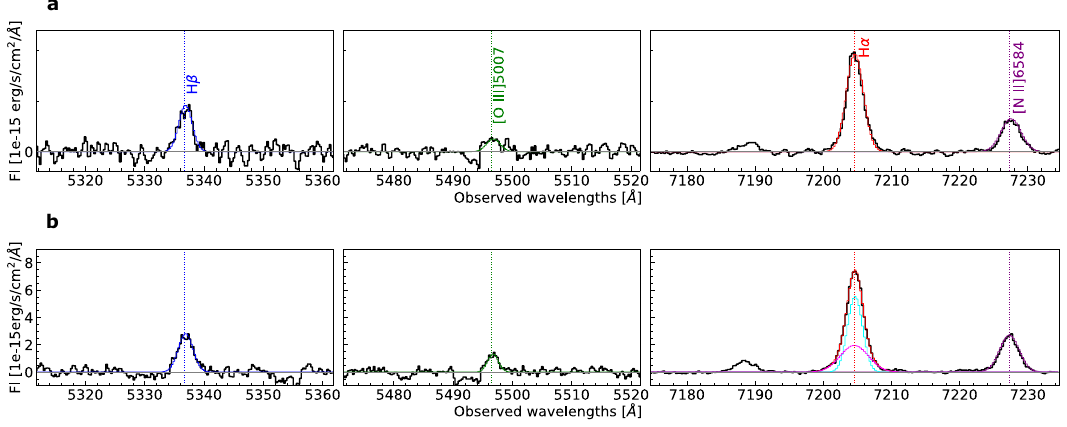}
    
     \caption{\textbf{Stacked GTC/MEGARA spectra}. From left to right, H$\beta$, [O III] 5007, and H$\alpha$ regions are presented. \textbf{a}, Stacked spectra from the six pixels in the FRB region (as shown in Extended Data Fig. \ref{fig:megara_halpha+ebv+lines}). \textbf{b}, Stacked spectra of the entire galaxy. Individual spectra were shifted to align with the reference redshift ($z=0.0978$) before stacking to compensate for any H$\alpha$-related
velocity effect. Colored vertical dotted lines mark the centroids of the three lines at the reference redshift.
We observe an absorption line at approximately $5493$ \AA. However, the nature of this spectral feature remains unrecognized by our analysis. This line is observed in each individual spectrum, too.
    \label{fig:MEGARA_total_spectrum}}
\end{EDfigure}

\clearpage
%%%%%%%%%%%%%%%%%%%%%%%%%%%%%%%%

\begin{EDfigure}
    \centering
    \includegraphics[]{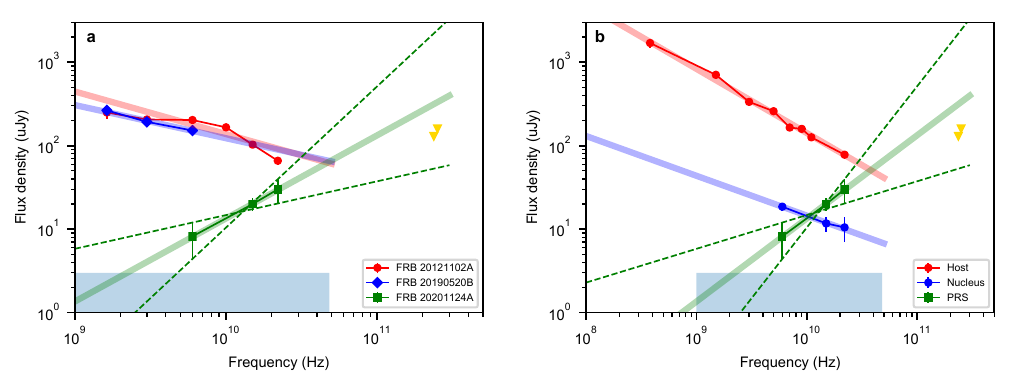}
    \caption{\textbf{The PRS spectrum.} \textbf{a}, radio spectra of the 3 PRSs known so far. \textbf{b}, Radio spectra of the PRS, the host galaxy's nuclear region, and the total host galaxy's emission. In both figures: power-law fits are reported as solid line; the green dashed lines indicate the maximum and minimum slope consistent with measurements within errors for the PRS presented in this work; the yellow triangles represent the NOEMA upper limits for \frb; the blue shaded area represents the region not detectable by the VLA, in the range 1--50 GHz (we assume a representative RMS = 1 $\mu$Jy beam$^{-1}$, reachable in $\sim$10h at 6 GHz).\label{fig:radio-SED}}
\end{EDfigure}

\clearpage
%%%%%%%%%%%%%%%%%%%%%%%%%%%%%%%%

\begin{SItable}
\caption{\textbf{Journal of radio and GTC/MEGARA observations.} The estimated flux densities for the PRS are shown, too.\label{tab:vla_journal}
}

    \centering
    \scalebox{0.9}{
    \begin{tabular}{cccccccc}
        \hline
        Telescope   & Date & Frequency & Bandwidth & FWHM & P.A. & RMS & Flux density \\
                    & dd/mm/yyyy & (GHz) & (GHz) & (arcsec) & (deg) & ($\mu$Jy/beam) & ($\mu$Jy) \\ 
        \hline
        VLA$^\dagger$  & 06/03/2022 & 6  & 4 & $0.38 \times 0.36$ & $-$51 & 2.0 & 8.2$\pm$3.8 \\
        VLA         & 07/10/2021    & 15 & 6 & $0.38 \times 0.36$ & +3.6 & 2.3 & 20.0$\pm$3.5 \\
        VLA         & 03-07/01/2022 & 22 & 8 & $0.37 \times 0.32$ & $-$53 & 3.0 & 30.0$\pm$9.7 \\
        NOEMA &  12/03/2023    & 236.5 & 7.6& $0.7 \times 0.42$ & +15 & 43 & $<130^\ddagger$ \\
        NOEMA &  12/03/2023    &  250  & 7.6 & $0.66 \times 0.39$ & +15 & 53 & $<160^\ddagger$ \\
%        GTC/MEGARA  & 28/10/2022 & 4.5e05 & 0.8e-05 & 0.62 &  \\
        \hline
    \multicolumn{8}{l}{
      \footnotesize{{\bf Notes:} $^\dagger$ re-imaged archival data from \cite{2024ApJ...961...44D}; $^\ddagger$ 3$\sigma$ upper limits}
    }
    \end{tabular}
    }

\vspace{1cm}

  \scalebox{0.9}{    
  \begin{tabular}{cccccc}          
   \hline
    GTC/MEGARA$^+$ & Date & Band & Exposure & R.L.D.* & R$^\dagger$ \\  % table heading
               &      & (\AA) & (s) & (\AA\ px$^{-1}$) & \\
    \hline    
       VPH LR-V & 28/10/2022 &  5165.6 -- 6176.2 & $3 \times 1800$ & 0.27 & 6078 \\
       VPH LR-R & 28/10/2022 &  6158.3 -- 7287.7 & $3 \times 1200$ & 0.31 & 6100 \\
    \hline
    \multicolumn{6}{l}{\footnotesize {\bf Notes:} $^+$ See the \hyperlink{https://www.gtc.iac.es/instruments/megara/media/MEGARA_cookbook_1I.pdf}{instrument cookbook}; * Reciprocal linear dispersion; $^\dagger$ Resolving power ($\lambda/\Delta\lambda_{\rm FWHM}$).}
\end{tabular}
}

\end{SItable}

\clearpage
%%%%%%%%%%%%%%%%%%%%%%%%%%%%%%%%%%%%%%%%%%%%%%%%%%%%%%%%%%%%%%%%

\begin{SItable}
\caption{\textbf{FRB sample considered in this work.} RM measurements (or upper limits) and flux densities of the PRS are reported.\label{tab:tab1}}

\scalebox{0.65}{
\begin{tabular}{ccccccccccc}
    \hline
FRB Name  &  ${\rm DM_{obs}}^{\rm a}$   & $\rm{DM_{MW}}^{\rm b}$   & $z^{\rm c}$    &   $d_{\rm{L}}^{\rm d}$  & ${\rm RM}^{\rm e}$ & $F_{\nu}^{\rm f}$\ & $\nu^{\rm g}$ & $L_{\nu}^{\rm h}$  & References\\
&  ($\unit{pc\,cm^{-3}}$)   &  ($\unit{pc\,cm^{-3}}$)   &     &   ($\unit{Gpc}$)  &  ($\unit{rad\,m^{-2}}$)  & ($\mu\unit{Jy}$)   &  ($\unit{GHz}$) & ($10^{29}\unit{erg\,s^{-1}Hz^{-1}}$)  &  \\
    \hline
FRB  20121102A &  $557$       & $188 $ & $0.19273$    &   $0.98$  &  $1.4\times10^{5}$  &   180  & $1.7$ & $2.1$  & 1,2,3,4\\

FRB 20180916B  &  $348.76$ & $200$& $0.0337$&   $0.15$  & $-114.6$   & $<18$    & $1.6$  & $<0.0048$   & 5,6\\

FRB 20180924B  &  $361.42$ & $40.5$ & $0.3214$    &   $1.74$  & $14$         & $<20$     & $6.5$ & $<0.72$ & 7\\

FRB 20181112A	&  $589.27$ & $102$ & $0.47550$  & $2.76$ & $10.9$ & $<21$ & $6.5$ & $<1.91$ & 8\\

FRB 20190520B	&  $1204.7$ & $113$ & $0.241$  & $1.25$ & $-3.6\times10^{4}$ & $202$ & $3$ & $3.8$ & 9,10\\

FRB 20200120E   &   $87.8$ &   $55$ & $0.0008$ & $0.0036$ & $-36.9$ & $<20$ & $1.5$ & $<3.1\times10^{-6}$ & 11,12\\

FRB 20201124A	&  $413$ & $140$ & $0.0978$  & $0.47$ & $-889.5$ & $20$ & $15$ & $0.053$ & 13, this work\\

FRB 20210117A	&  $729.1$ & $34$ & $0.214$ & $1.10$ & $43$  & $<10$ & $6$ & $<0.15$ & 14\\

FRB 20220912A   & $220$ & $125$ & $0.0771$ & $0.36$ & $-0.08$ & $<48$ & $1.4$ & $<0.074$ & 15,16\\

\hline

FRB 20110523A  &  $623.3$ & $43.52$ & $0.58^{+0.21}_{-0.21}$ & $3.5^{+1.6}_{-1.4}$ & $-186.1$ &  $<40$ & $0.8$ & $<5.8^{+6.6}_{-3.7}$ & 17\\

FRB 20150215A  &  $1105.6$  & $427.2$& $0.69^{+0.22}_{-0.22}$    &   $4.3^{+1.7}_{-1.6}$  & $1.5$   & $<6.48$   & $10.1$ & $<1.4^{+1.4}_{-0.9}$  & 18\\

FRB 20150418A	&	$776.2$ & $188.5$ & $0.59^{+0.21}_{-0.21}$ & $3.6^{+1.6}_{-1.4}$ & $36$  & $<70$ & $1.4$ & $<11^{+12}_{-7}$ & 19\\

FRB 20150807A	&	$266.5$ & $36.9$ & $0.17^{+0.10}_{-0.11}$ & $0.85^{+0.57}_{-0.57}$ & $12$  & $<240$ & $5.5$ & $<2.1^{+3.7}_{-1.8}$ & 20\\

FRB 20160102A  &  $2596.1$  & $13$   & $3.04^{+0.51}_{-0.48}$     &   $26.4^{+5.4}_{-4.9}$ & $-220.6$   & $<30$     & $5.9$ & $<249^{+112}_{-84}$   & 21,22\\

FRB 20180309A  &  $263.42$ & $44.69$& $0.16^{+0.10}_{-0.10}$    &   $0.79^{+0.57}_{-0.51}$  & $<150$     & $<105$     & $2.1$ & $<0.8^{+1.5}_{-0.7}$  & 23\\

FRB 20191108A	&  $588.1$ & $52$ & $0.53^{+0.20}_{-0.20}$ & $3.1^{+1.5}_{-1.3}$ & $474$  & $<213$ & $1.4$ & $<24^{+29}_{-16}$ & 24\\

   \hline	
   \end{tabular}
}
    {\flushleft
    \footnotesize{{\bf Notes:} $^{\mathrm{a}}$The observed dispersion measures (DMs) of FRBs.
$^{\mathrm{b}}$The Galactic DM contributions from the references or the NE2001 from FRB catalog \cite{pet16}.
$^{\mathrm{c}}$The measured/inferred redshifts of FRBs. 
The redshifts in the top row are directly measured by the host galaxies of FRB. The redshifts in the bottom row are inferred by the extragalactic DMs, see \cite{yang20}. 
The $\Lambda$CDM cosmological parameters are taken as $\Omega_{\rm m}=0.315$, $\Omega_bh^2=0.02237$, and $H_0=67.36\,{\rm km\,s^{-1}\,Mpc^{-1}}$ \cite{2020A&A...641A...6P}.
$^{\mathrm{d}}$ The luminosity distance calculated by redshift.
$^{\mathrm{e}}$ The observed RMs of FRBs. For the repeating sources, we take the largest absolute value for the observed RMs.
$^{\mathrm{f}}$ The observed PRS flux densities. The upper limits here are corrected to the $3\sigma$ flux density limits.
$^{\mathrm{g}}$ The measured frequencies of the persistent radio source.
$^{\mathrm{h}}$ The specific luminosities of the PRS inferred by the observed flux density and distance.
References: (1) \cite{spi14}; (2) \cite{ten17}; (3) \cite{mar17}; (4) \cite{mic18}; (5) \cite{chi19}; (6) \cite{mar20}; (7) \cite{ban19}; (8) \cite{pro19}; (9) \cite{niu22}; (10) \cite{ann23}; (11) \cite{Bhardwaj21}; (12) \cite{Kirsten22} (13) \cite{xu22}; (14) \cite{bha23}; (15) \cite{Zhang23}; (16) \cite{Hewitt23}; (17) \cite{mas15};
(18) \cite{pet17}; (19) \cite{kea16}; (20) \cite{rav16}; (21) \cite{bha18}; (22) \cite{cal18}; (23) \cite{osl19}; (24) \cite{con20}. 
}}
\end{SItable}

\clearpage

\begin{SItable}
\caption{\textbf{Emission lines from the stacked spectra.} Fluxes and uncertainties are in units of [erg/s/cm$^2$]. Fluxes are corrected for Galactic extinction only.\label{tab:megara_stacked}}

%whereas dispersion velocities $\sigma$ are in units of [km/s] and are not corrected for instrumental effects. }
\centering
\begin{tabular}{lccc}
\hline
Line & Param. & FRB reg. &  Global  \\
\hline

H$\alpha$ (1st comp.) & flux & $3.03\times10^{-15}$ & $1.53\times10^{-14}$ \\
 & $\sigma$ & $2.97\times10^{-17}$ & $1.01\times10^{-16}$ \\
 %& $\sigma$ & $52.4$ & $45.9$\\
 \hline
 H$\alpha$ (2nd comp.) & flux &  & $1.01\times10^{-14}$ \\
 & $\sigma$ && $1.34\times10^{-16}$ \\
 \hline
H$\beta$ & flux & $1.19\times10^{-15}$ & $7.96\times10^{-15}$ \\
 & $\sigma$ & $7.68\times10^{-17}$ & $4.12\times10^{-16}$ \\
 \hline
[N II] 6584 & flux & $1.20\times10^{-15}$ & $9.69\times10^{-15}$ \\
 & $\sigma$ & $3.1\times10^{-17}$ & $1.67\times10^{-16}$ \\
 \hline
[O III] 5007 & flux & $5.63\times10^{-16}$ & $3.9\times10^{-15}$ \\
& $\sigma$ & $5.03\times10^{-16}$ & $3.17\times10^{-16}$ \\

\hline
\end{tabular}
\end{SItable}

\clearpage
\section*{SUPPLEMENTARY METHODS}\label{sec_SM}
\vspace{.5cm}

\noindent\textbf{Host galaxy characterization with GTC/MEGARA}
\\
The continuum emission of the projected galaxy extends approximately 9 kpc along the major axis ($a$) and around 5 kpc along the minor axis ($b$), without accounting for PSF effects. 
By assuming a straightforward circular galactic geometry, we calculated an inclination angle $i$ of approximately 56\fdg3 based on the ratio of $a/b$.

\begin{figure}
    \centering
    \includegraphics{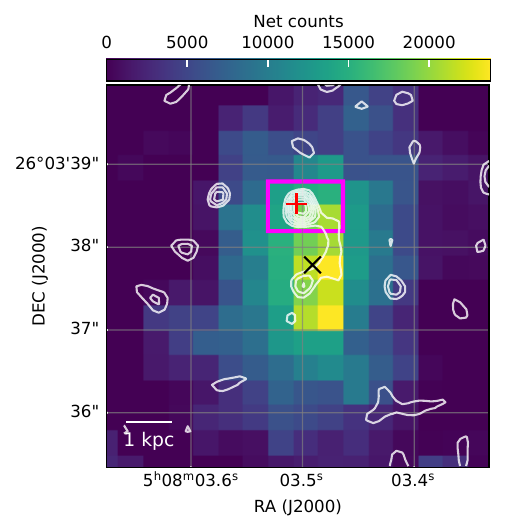}
    \caption{\textbf{GTC/MEGARA map of the net counts of the galaxy.} 
    The white contours represent the VLA image at 15 GHz, whereas the red and black crosses represent the FRB and galactic centre, respectively.
    The magenta rectangle defines the galactic region encompassing six adjacent MEGARA pixels surrounding the FRB and PRS zone.\label{fig:megara_counts}}
\end{figure}

\begin{figure}
    \centering
    \includegraphics[width=12cm]{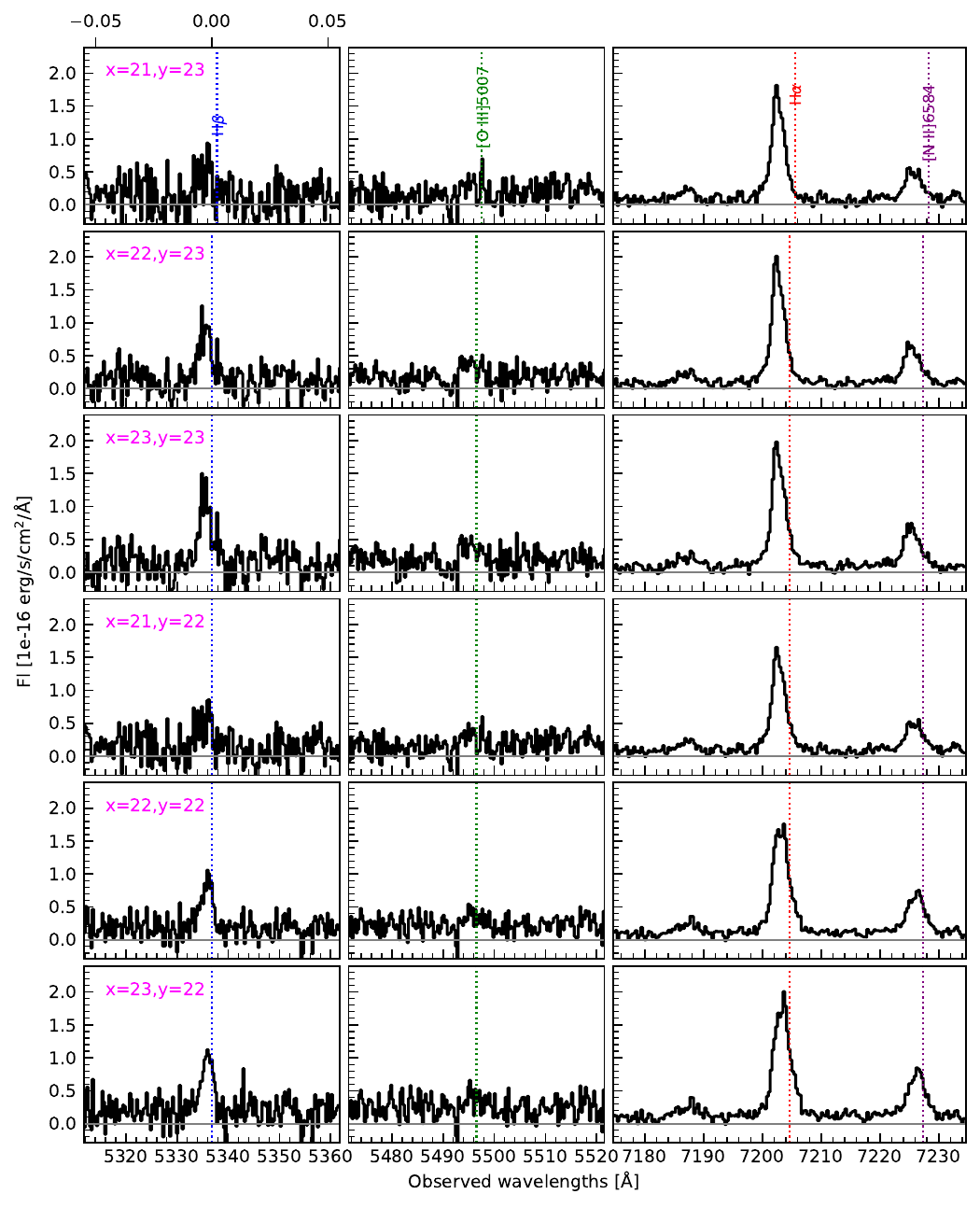}
    \caption{\textbf{Compilation of R+V grisms GTC/MEGARA spectra}. From left to right: the H$\beta$, [O III] 5007, and H$\alpha$ regions, respectively. The spectra in this figure are arranged to represent the spectral features of pixels within a rectangular selection that defines the FRB region (see the magenta box in Fig. \protect\ref{fig:megara_counts}).  The first three spectra from the top correspond to the first pixel row of the FRB region, while the last three spectra represent the second pixel row.
   The colored vertical dotted lines indicate the spectral positions of the lines at the reference redshift ($z=0.0978$), revealing that the FRB region is blue-shifted by 60--80 km s$^{-1}$ relative to the reference redshift (see Fig. \ref{fig:megara_velocities}).
   We observe an absorption line at approximately 5493 \AA\ across all spaxels in the region; however, the nature of this spectral feature remains unrecognized by our analysis. Note that the H$\beta$ and [O III] 5007 spectral regions have a pixel dispersion of 0.27 \AA, corresponding to approximately $15.2$ and $14.7$ km/s/pixel, respectively. Meanwhile, the H$\alpha$ region has a pixel dispersion of 0.31 \AA\, corresponding to approximately $12.9$ km s$^{-1}$ pixel$^{-1}$.
    \label{fig:MEGARA_FRB_spectra}}
\end{figure}

In Fig. \ref{fig:megara_counts}, we present the map of the net counts of the galaxy. 
Figure \ref{fig:MEGARA_FRB_spectra} shows a compilation of optical spectra of the galaxy. In particular, we show the spectra of H$\beta$, [O III] 5007 and H$\alpha$ corrected for Galactic extinction for the six pixels surrounding the FRB and PRS (referred to as the FRB region), from left to right. The FRB region is delineated by a magenta rectangle in Fig. \ref{fig:megara_counts}.
These spectra show blue-shifted emission lines with respect to the reference redshift ($z=0.0978$), indicating rotational velocities between $-60$ and $-80$ km/s, as shown in Fig. \ref{fig:megara_velocities}. 
The H$\alpha$ line in the right panels of Fig. \ref{fig:MEGARA_FRB_spectra} shows a clear and highly significant detection.
Despite its intrinsic weakness compared to H$\alpha$ by a factor of about 3, H$\beta$ performs quite well, even in a spectral region with a RMS roughly 2.7 times worse than that in the H$\alpha$ spectral region. 
Indeed, the standard deviation of the continuum is about $1.7 \times 10^{-17}$ erg/s/cm$^2$/\AA\ for H$\beta$ compared to $6.2 \times 10^{-18}$ erg/s/cm$^2$/\AA\ for H$\alpha$.
Conversely, the [O III] 5007 line shown in the middle panels of Fig. \ref{fig:MEGARA_FRB_spectra} is either absent or very weak in all spectra. We will analyse this spectral feature in more detail later in this section.
We have identified an unexpected absorption feature, consistently appearing at an observed wavelength of approximately $\lambda\sim5493$\;\AA\ in all six pixels within the FRB region. We anticipate that this feature is also present in the stacked spectrum of the entire galaxy which we will introduce later in this session.

We obtained measurements of the emission lines H$\alpha$, H$\beta$, [O III] 5007, and [NII] 6584 at the spaxel level through the following procedure:
as a first step we corrected each spectrum for Galactic extinction (E(B$-$V) = 0.652) using the \cite{Cardelli1989} extinction law. Subsequently, we removed the underlying stellar continuum by subtracting a pseudo-continuum, defined as a straight line connecting adjacent bluer and redder regions to the emission component, devoid of other relevant spectral features. 
We used a Python-Astropy routine based on the Levenberg-Marquardt algorithm and least squares statistic \cite{Astropy2022} to fit up to two Gaussian components to each line, selecting as the best-fit solution the one that minimized the reduced chi-square statistic toward 1.
We assessed the validity of our Gaussian fits by directly integrating the total line flux within a wavelength interval centered in the line centroid and spanning twice the standard deviation of the best-fitted Gaussian for each line.
Finally, we used a bootstrapping technique to quantify uncertainties in the Gaussian parameters for each emission line (i.e., amplitude, line's centroid, and standard deviation $\sigma$). We iteratively sampled and fitted the spectra $1000$ times, introducing Gaussian noise at a spectral pixel level with a standard deviation equal to the RMS of the spectrum within the spectral regions corresponding to the pseudo-continuum of the respective line. This bootstrapping process yields a Gaussian distribution of parameter values, with the standard deviation providing a robust estimate of the uncertainty associated with the fitted parameter.

In conjunction with our pixel-level analysis of the galaxy, this study includes two additional spectra, shown in Extended Data Fig. 4. In panel \textbf{a} we show the composite spectrum derived by stacking the R and V grism spectra of the six pixels defining the FRB region previously described in Fig. \ref{fig:megara_counts}.
The individual spectra in this composite spectrum have been corrected to a reference redshift of $z = 0.0978$. This correction mitigates any radial velocity shift that would broaden the width of spectral lines such as H$\alpha$ and other fainter lines. In this way, we prevent possible deviations from the Gaussian shape of the lines and limit the dilution of the flux of fainter features such as H$\beta$ across multiple spectral elements, which greatly improves the detectability of such faint features, even if this results in the loss of information about the kinematics of the gas in the region.
Note that the kinematics and dynamics of the gas in the galaxy will be thoroughly discussed later in this paper.
In addition, in Extended Data Fig. 4b we present the global spectrum of the whole galaxy obtained by stacking the R and V spectra, corrected to $z = 0.0978$, of all pixels within the rectangular region delimited by RA 05:08:3.4$\div3.6$~s and Dec 26:03:35$\div40$~arcsec. The line fluxes and uncertainties for both stacked spectra are listed in Extended Data Tab. 3. 
%
%\begin{table}
%\caption{Emission lines in the stacked spectra. Fluxes and uncertainties are in units of [erg/s/cm$^2$]. Fluxes are corrected for Galactic extinction only.}\label{tab:megara_stacked}
%whereas dispersion velocities $\sigma$ are in units of [km/s] and are not corrected for instrumental effects. }
%\centering
%\begin{tabular}{lccc}
%\hline
%Line & Param. & FRB reg. &  Global  \\
%\hline
%
%H$\alpha$ (1st comp.) & flux & $3.03\times10^{-15}$ & $1.53\times10^{-14}$ \\
% & $\sigma$ & $2.97\times10^{-17}$ & $1.01\times10^{-16}$ \\
% %& $\sigma$ & $52.4$ & $45.9$\\
% \hline
% H$\alpha$ (2nd comp.) & flux &  & $1.01\times10^{-14}$ \\
% & $\sigma$ && $1.34\times10^{-16}$ \\
% \hline
%H$\beta$ & flux & $1.19\times10^{-15}$ & $7.96\times10^{-15}$ \\
% & $\sigma$ & $7.68\times10^{-17}$ & $4.12\times10^{-16}$ \\
% \hline
%[N II] 6584 & flux & $1.20\times10^{-15}$ & $9.69\times10^{-15}$ \\
% & $\sigma$ & $3.1\times10^{-17}$ & $1.67\times10^{-16}$ \\
% \hline
%[O III] 5007 & flux & $5.63\times10^{-16}$ & $3.9\times10^{-15}$ \\
%& $\sigma$ & $5.03\times10^{-16}$ & $3.17\times10^{-16}$ \\
%
%\hline
%\end{tabular}
%\end{table}
%
The motivation behind the creation of these stacked spectra is to increase the RMS, thereby improving the statistical significance of weak spectral lines, including the H$\beta$ and [O III] 5007 lines.

%This allows us to compare our results with those of previous studies.

%%%%%%%%%%%%%%%%%%%%%%%%%%%%%%%%
%\begin{figure}
%    \centering
%    %\includegraphics[width=7cm]{Lines_v1.pdf}
%    \includegraphics{Figures/Lines_v1_2cols.pdf}
%    \caption{GTC/MEGARA: Maps of the emission lines [N II] 6584 (a), H$\beta$ (b), and [O III] 5007 (c). 
%    Pixels with an H$\alpha$ significance below a $3\sigma$ threshold are omitted. In cases where the line significance falls below a $2\sigma$ we give a line upper limit, and represent these pixels in grey-scale. Note that the tick labels above the two colorbars represent upper flux limits and actual flux values for grayscale and coloured pixels, respectively. The red and black crosses represent the FRB and galactic centre, respectively.
%    The magenta rectangle define the galactic region encompassing six adjacent MEGARA pixels surrounding the FRB and PRS zone.\label{fig:megara_lines}}
%\end{figure}

%%%%%%%%%%%%%%%%%%%%%%%%%%%%%%%%%%%%

%%%%%%%%%%%%%%%%%%%%%%%%%%%%%%%%%%%%

%%%%%%%%%%%%%%%%%%%%%%%%%%%%%%%%%%%%

\noindent\textbf{Emission line maps}
\\
Extended Data Fig. 3 shows the galaxy's H$\alpha$ emission line map, which has been corrected for Galactic extinction. In our analysis, we need reliable measures of the H$\alpha$ line, hence we exclude pixels whose H$\alpha$ significance is below $3\sigma$. In this map, only pixels with an H$\alpha$ significance above $3\sigma$ are shown. This H$\alpha$ map serves as the starting point for our analysis of the GTC/MEGARA optical spectra, enabling us to derive crucial physical properties of the gas phase in the galaxy, including gas kinematics and dynamics, ionization levels, SFR, and gas-phase metallicity.

Extended Data Fig. 3 shows maps of three emission lines within the wavelength range covered by our GTC/MEGARA data, specifically, [N II] 6584, H$\beta$, and [O III] 5007 lines, from top to bottom. 
Each panel in Extended Data Fig. 3 uses two color schemes: one represents pixels where the line is detected above $2\sigma$ (colored scale pixels), and the other represents pixels with flux upper limits.
We determined upper limits for the lines at a $2\sigma$ significance level for all pixels where the signal-to-noise ratio of the line fell below $2$. Here, $\sigma$ represents the standard deviation of the Gaussian best-fit of the stacked spectrum of the whole galaxy, as shown in Extended Data Fig. 4b.
The [N II] 6584 and H$\beta$ line maps exhibit similar flux distribution patterns as observed in the H$\alpha$ map, albeit with fainter emissions towards the galaxy's outer regions (although H$\beta$ is intrinsically fainter, approximately three times less intense than H$\alpha$). 
The [O III] 5007 line, instead, remains undetected at a $2\sigma$ level across nearly all pixels of the galaxy, although it is detected in the global stacking, as also shown in Extended Data Fig. 3 and Fig. 4. This faint emission in the [O III] lines may be attributed to either high gas metallicity or a low ionization state resulting from diminished UV radiation \cite{Citro2017}.
We will explore this ionization-metallicity degeneracy in the next subsection.

%%%%%%%%%%%%%%%%%%%%%%%%%%%%%%%%%%%%%%%%%%%%
\noindent\textbf{Derived quantities from optical spectra}\\
%%%%%%%%%%%%%%%%%%%%%%%%%%%%%%%%%%%%
\begin{figure}
    \centering
    \includegraphics{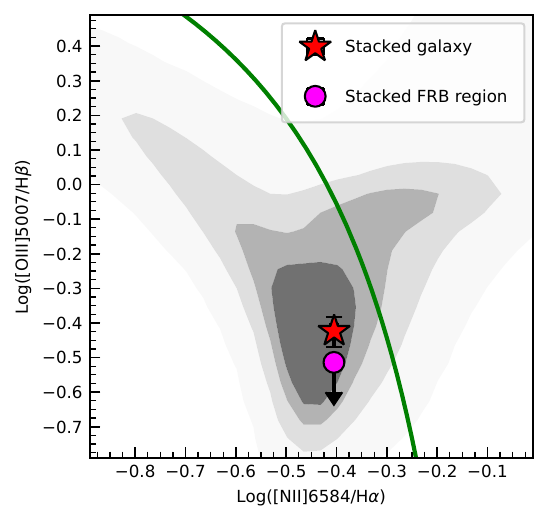}
    \caption{\textbf{The BPT diagram \cite{BPT1981} correlating log([O III] 5007/H$\beta$) and log([N II] 6584/H$\alpha$).} The red star marks the position of the entire galaxy's stacked spectrum, while the magenta circle represents upper limit position of the FRB region's stacked spectrum. Error bars represent $1\sigma$ standard errors for the emission line ratios. The gray shade areas represent the distribution of SDSS galaxies as a reference for the local Universe. The green curve, derived from \cite{Kauffmann2003}, serves as an empirical criterion to distinguish ionization sources, with both data points lying below it. Our data demonstrates that the ionization source in the galaxy is predominantly radiation from massive and young stars.\label{fig:megara_BPT}}
\end{figure}

In the previous section, we established that the [O III] emission is lacking in the FRB region and generally weak and undetectable throughout the entire galaxy. This intriguingly suggests recent star formation quenching. Following a sharp decline in star formation, the high-energy UV radiation emitted by short-lived, massive O-type stars decreases significantly within a few million years.
This would lead to a rapid decline in the luminosity of lines with high ionisation, such as the [O III] emission lines. However, emission lines associated with lower ionisation energies, such as the Balmer lines, would continue to be detectable as long as late O and early B-type stars with longer lifetimes remain in the main sequence phase.
However, [O III] 5007 emission can be depressed in a similar way also by high gas metallicity, leading to the so-called ionization-metallicity degeneracy \cite{Citro2017}. 
 
The ionization-metallicity degeneracy on the [O III] emission can be mitigated by using pairs of emission line ratios, each related to either metallicity or ionization. For example, the [N II] 6584/[O II] 3727 or [S II]/H$\alpha$ ratios can serve as proxies for metallicity, whereas [O III] 5007/H$\alpha$ or [Ne III] 3869/[O II] 3727 ratios can serve as proxies for ionization levels \cite{Citro2017, Quai2018, Quai2019, Dhiwar2023}.
Our GTC/MEGARA wavelength coverage is limited to the four aforementioned lines, thus limiting us to disentangling the two underlying physical properties responsible for the faintness of the [O III] 5007 line. 
Nevertheless, the four aforementioned emission lines allow us to explore the diagnostic diagram introduced by \cite{BPT1981}, hereafter referred to as BPT, which relates log([O III] 5007/H$\beta$ and log([N II] 6584/H$\alpha$ to the ionization level and metallicity contempt in the galaxy.

The location of a galaxy (or galactic region) in the BPT plane is very sensitive to the source of UV radiation that can be radiation from the hot stellar photospheres of recently formed massive stars, AGN activity, shock processes, or combinations of these.
Figure \ref{fig:megara_BPT} shows the BPT diagram of the galaxy for the two stacked spectra of the FRB region and the whole galaxy, as well as the distribution of the SDSS-DR8  \cite{Eisenstein2011} Main galaxies as a reference for low redshift galaxies (gray shaded area).
We have adopted the criterion of \cite{Kauffmann2003} to distinguish between ionisation due to star formation (i.e. data points lying below the green curve in Fig. \ref{fig:megara_BPT}), and other radiation sources (e.g., ionization due to active galactic nuclei). 
According to the \cite{Kauffmann2003}, both the stacked FRB spectrum and the entire galaxy have radiation dominated by star formation.

%%%%%%%%%%%%%%%%%%%%%%%%%%%%%%%%

%\subsubsection*{The gas phase metallicity}

We estimated the gas phase metallicity of the galaxy using the strong line metallicity indicator O$_3$N$_2$, as initially defined by \cite{Alloin1979} as log([O III] 5007/H$\beta$/[N II] 6584/H$\alpha$). 
We adopted the metallicity calibration by \cite{Curti2017} with a solar abundance value of 12+log(O/H)$_\odot = 8.69$ \cite{AllendePrieto2001}.
We measured slightly super solar values of 12+log(O/H) $=8.75$ in the stacked spectra of the entire galaxy, and a lower limit of $8.77$ in the FRB region. 
To mitigate the ionization-metallicity degeneracy in the [O III] line, we used predictions from the \cite{Citro2017} photoionization models using the four emission lines defining the BPT diagram. 
We found that the stacked spectrum of the entire galaxy is compatible with emission from HII regions with ionization level U~$\sim -3.4$, typical of a star-forming galaxy, and solar metallicity. 
For the FRB region, we obtained a metallicity compatible with solar metallicity, and an upper limit in the ionization level U~$\leq -3.5$. 
The upper limit in the parameter U is due to the low significance level of the [O III] line of the stacked spectrum in the FRB region. 
It is worth noting that the upper limit on the parameter U arises from the lower significance level of the [O III] line in the stacked spectrum of the FRB region. Consequently, considering the current depth and S/N of our optical data, along with the absence of information from other emission lines capable of independently sensing changes in the ionization parameter and gas phase metallicity, we are unable to resolve the ionization-metallicity degeneracy conclusively. Thus, we cannot confirm or exclude the possibility that the FRB region has recently experienced a decline in star formation.

%
%
%%%%%%%%%%%%%%%%%%%%%%%%%%%%%%%%%%%%

\begin{figure}
    \centering
    \includegraphics[width=15cm]
    {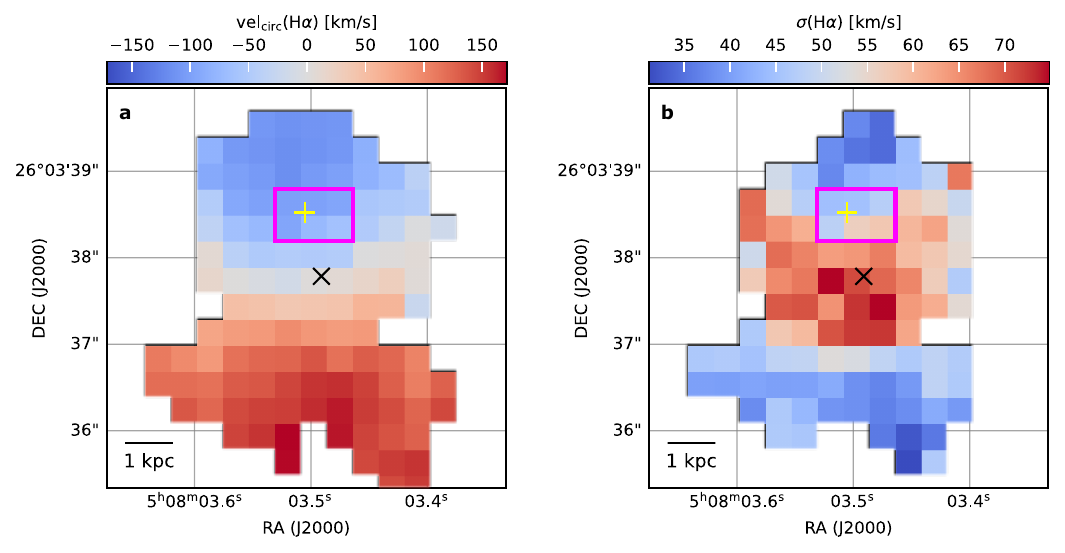}
    \caption{\textbf{GTC/MEGARA}. \textbf{a},  Circular velocity and \textbf{b}, velocity dispersion corrected for instrumental resolution maps derived from H$\alpha$ emission.
    \label{fig:megara_velocities}}
\end{figure}

%%%%%%%%%%%%%%%
\noindent\textbf{Kinematics and dynamics of the gas-phase}\\
\label{sec:kinematics}
Figure \ref{fig:megara_velocities}a shows the circular velocity map ($\text{vel}_\text{circ}$) for the galaxy. We derived this map by de-projecting the H$\alpha$ radial velocity ($\text{vel}_\text{los}$), corrected for the relative velocity at the reference redshift (z=0.0978), onto the galactic plane using the equation:

\begin{equation}
\text{vel}_\text{circ} = \frac{\text{vel}_\text{los}}{\sin(i)},
\end{equation}

\noindent where $i=56.3^\circ$ represents the inclination angle of the galaxy, which we estimated based on the net-counts map (Figure 1).
The resulting velocity map demonstrates that the galaxy exhibits well-ordered gas kinematics predominantly characterized by circular orbits. Rotation velocities range from 0 to $-100$ km/s (at the galactic center and a radial distance of $\sim3$ kpc, respectively) in the northern region, and from $0$ to approximately 150 km/s (at a radial distance of $\sim3.5$ kpc) in the southern region.

Figure \ref{fig:megara_velocities}b shows the H$\alpha$ velocity dispersion map ($\sigma(\text{H}\alpha)$) for the galaxy. To account for instrumental resolution effects, we subtracted in quadrature the sigma value (in units of \AA) derived from the Gaussian fit of the strong Earth sky line at 7276.405\;\AA\ from that of the measured H$\alpha$. With this approach we obtain a reliable correction as the reference sky line is nearby to the observed H$\alpha$ line and clearly detected.
The galaxy shows dynamics typical of a local star-forming galaxy \cite{Bundy2015} with velocity dispersion values reaching up to 80 km/s in the center and lower values of only 10--20 km/s in the outskirts.

We report here a summary of our findings regarding the GTC/MEGARA data. 
We found a star-forming galaxy with an inclination angle of approximately $56^\circ$. The H$\alpha$ velocity map reveals predominantly circular gas kinematics with velocities ranging from $-100$ km/s (north) to 150 km/s (south). Emission line measurements, including H$\alpha$, H$\beta$, [O III] 5007, and [NII] 6584, corrected for extinction, provide insights into the gas metallicity and star formation rates (SFR). Gas metallicity is slightly super-solar, and the BPT diagram indicates radiation dominated by star formation. The SFR map shows a peak near the galactic center as identified by the HST images, with $\Sigma\;$SFR of $\sim 0.32$ to $0.64$ M$_\odot$/yr/kpc$^2$. The FRB region exhibits lower SFR levels, and analysis of stacked spectra yields an estimated SFR of $0.41^{+0.02}_{-0.01}$ M$_\odot$/yr, significantly lower than SFR obtained from the radio emissions.

%%%%%%%%%%%%%%%%%%%%%%%%%%%%%%%%%%%%

\noindent\textbf{Host galaxy SED Analysis}\\
\label{host}
We performed a re-analysis of the host galaxy SED to include the recently acquired NOEMA data. The NOEMA limits set a stringent upper bound to the flux in the mm range, and are over predicted by the model presented in our previous work\cite{2021A&A...656L..15P}. We note that these limits seem to be marginally consistent with the best-fit model presented by ref.\cite{Fong21}, while ref.\cite{Ravi21} does not present their SED out to these wavelengths for comparison.

In order to determine how the NOEMA data at far-infrared wavelengths (FIR) changes the results, we re-modeled the SED using \textsc{prospector}\cite{Johnson2019} with the methods utilized in ref.\cite{2021A&A...656L..15P}. We similarly include data spanning the $ugrizJHK$ bands, and in this case we include all four WISE bands ($W1$, $W2$, $W3$, $W4$) in order to properly constrain the mid-infrared (MIR) SED.
In order to constrain the shape of the MIR SED, we likewise included the contribution of dust emission from polycyclic aromatic hydrocarbons (PAHs), which uses the model presented by \cite{Draine2007}. 
We additionally test for the presence of an AGN contributing to the MIR. These effects were not included in ref.\cite{2021A&A...656L..15P}.  

The dust emission model has three free parameters: the fraction of dust mass in PAHs $Q_\textrm{PAH}$, lower limit to the strength of starlight in units of the Milky Way $U_\textrm{min}$, and the fraction $\gamma_e$ of dust irradiated by starlight stronger than $U_\textrm{min}$, where the power-law distribution of starlight is a power-law with slope $\alpha=2$. 
We adopted broad uniform priors for these parameters: $0.1<Q_\textrm{PAH}<10.0$, $0.1<U_\textrm{min}<25$ (bounded by the wavelengths included in the fit), and $0.0<\gamma_e<0.15$, as outlined in \cite{Leja2018}. The AGN model utilizes the templates from \cite{Nenkova2008} and is parametrized by the fraction of bolometric stellar luminosity  $f_\textrm{AGN}$ emitted by the AGN, and the optical depth of the torus $\tau_\textrm{AGN}$. We utilize log uniform priors of $0<f_\textrm{AGN}<3$ and $5<\tau_\textrm{AGN}<150$, as in \cite{Leja2018}. 

\begin{figure}
\centering
\includegraphics[width=14cm]{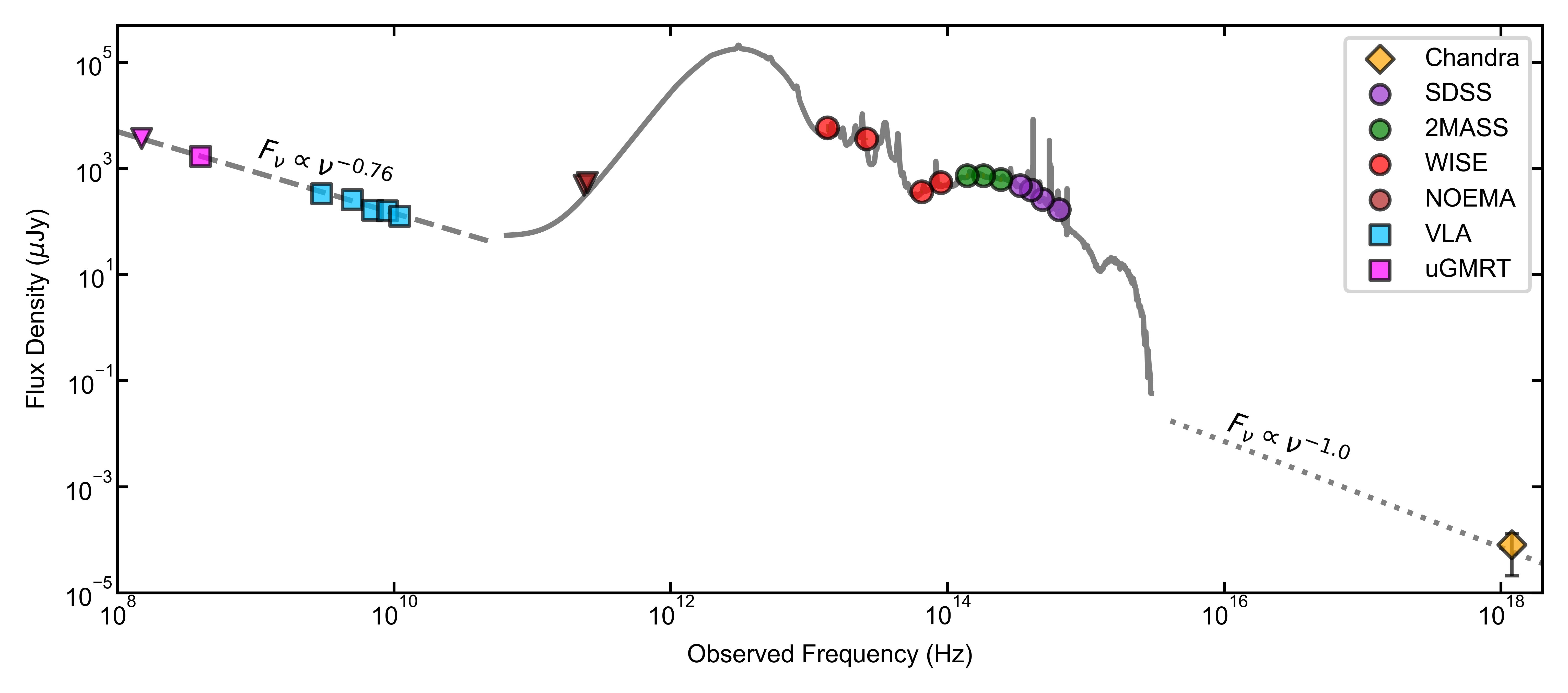}

    \caption{\textbf{SED of the host galaxy of FRB 20201124A from radio to X-rays.} The NOEMA limits are shown as dark brown  downward triangles. We have only included the VLA data obtained in the D configuration (reported in ref.\cite{2021A&A...656L..15P}).\label{fig:broadband_SED_wRadio}}
\end{figure}

The best-fit SED is shown in Figure \ref{fig:broadband_SED_wRadio}. We derive the following best-fit parameters: a significant intrinsic extinction $A_V=0.81^{+0.17}_{-0.14}$, a near solar metallicity $Z/Z_\odot=0.7\pm0.2$, a stellar mass $\log(M_*/M_\odot)=9.98^{+0.07}_{-0.08}$, an old mass-weighted stellar age $t_m=3.5\pm1.2$ Gyr, an e-folding time $\tau=4.1^{+2.9}_{-2.4}$, a star formation rate SFR = $2.4\pm0.3$ $M_\odot$ yr$^{-1}$. The dust parameters, in particular $Q_\textrm{PAH}=7\pm2\%$, are relatively unconstrained, besides strongly favoring $\gamma<0.25$ and a larger value of $U_\textrm{min}>1.5$ ($3\sigma$ confidence level).
The preference towards a higher value of $U_\textrm{min}=17^{+5}_{-9}$ is due to the fact that a larger value of irradiance leads to a higher dust temperature, pushing the peak of the SED away from the NOEMA limits. We further find that an AGN is likely only contributing at the $\sim1\%$ level of the total bolometric luminosity, and at most $<5\%$, in agreement with the previous value ($\sim4\%$) derived by \cite{Fong21}.

In comparison to ref.\cite{2021A&A...656L..15P}, we find a smaller SFR and an older stellar population, closer to the values reported by \cite{Fong21}, but still with a slightly smaller stellar mass. 
A similar analysis not including an AGN returns nearly identical parameters for $A_V$, $Z$, $M_*$, $t_m$, $\tau$, and SFR compared to ref.\cite{2021A&A...656L..15P} (i.e., consistent within $1\sigma$ errors), while only changing the dust parameters, similarly favoring a high $U_\textrm{min}\approx$ 20--25. 

Following \cite{Aniano2012}, we can convert the dust model parameters \cite{Draine2007} to a temperature, for better comparison to our analytical constraints on the NOEMA data. We find a dust temperature of $T_\textrm{dust}=29.5^{+1.3}_{-3.4}$ K. 
%http://dustpedia.astro.noa.gr/Content/tempFiles/cigale/DustPedia_CIGALE_read_me_file.pdf
%https://iopscience.iop.org/article/10.1088/0004-637X/756/2/138/pdf
%We further convert the output to a total dust mass in the galaxy of \textcolor{purple}{$M_\textrm{dust}=$} \cite{Draine2007,Aniano2012}. 
%We note that the analytical constraints derived previously from NOEMA data are based on the local limits to the PRS, while here we use the global limits of the entire galaxy. 
%https://arxiv.org/pdf/astro-ph/0608003.pdf
%https://www.aanda.org/articles/aa/pdf/2021/11/aa41678-21.pdf

%%%%%%%%%%%%%%%%%%%%%%%%%%%%%%%%

%%%%%%%%%%%%%%%%%%%%%%%%%%%%%%%%%%%%%%%%%%%%%%%%%%%%
% REFERENCES
%%%%%%%%%%%%%%%%%%%%%%%%%%%%%%%%%%%%%%%%%%%%%%%%%%%%
\section*{REFERENCES}
   \bibliography{FRB_biblio} % your 

%%%%%%%%%%%%%%%%%%%%%%%%%%%%%%%%%%%%%%%%%%%%%%%%%%%%%%%%%%%%%%%%%%%
%\end{linenumbers}
\end{document}